\documentclass[twocolumn,tighten,times,twocolappendix,numberedappendix]{aastex631}

\usepackage{amsmath}
\usepackage{float}
\usepackage{color}
\usepackage{url}
\usepackage{etoolbox}

\usepackage{graphicx}

\usepackage[flushleft]{threeparttable}

\hypersetup{breaklinks=true,colorlinks=true,linkcolor=blue,citecolor=blue,filecolor=blue,urlcolor=blue}
\makeatletter
\patchcmd{\NAT@citex}
  {\@citea\NAT@hyper@{%
     \NAT@nmfmt{\NAT@nm}%
     \hyper@natlinkbreak{\NAT@aysep\NAT@spacechar}{\@citeb\@extra@b@citeb}%
     \NAT@date}}
  {\@citea\NAT@nmfmt{\NAT@nm}%
   \NAT@aysep\NAT@spacechar\NAT@hyper@{\NAT@date}}{}{}

\patchcmd{\NAT@citex}
  {\@citea\NAT@hyper@{%
     \NAT@nmfmt{\NAT@nm}%
     \hyper@natlinkbreak{\NAT@spacechar\NAT@@open\if*#1*\else#1\NAT@spacechar\fi}%
       {\@citeb\@extra@b@citeb}%
     \NAT@date}}
  {\@citea\NAT@nmfmt{\NAT@nm}%
   \NAT@spacechar\NAT@@open\if*#1*\else#1\NAT@spacechar\fi\NAT@hyper@{\NAT@date}}
  {}{}
  
\usepackage{array}
\newcolumntype{C}[1]{>{\centering\let\newline\\\arraybackslash\hspace{0pt}}m{#1}}

\usepackage{xspace}

\def\aj{AJ}
\def\araa{ARA\&A}
\def\apj{ApJ}
\def\apjl{ApJ}
\def\apjs{ApJS}
\def\ao{Appl.~Opt.}

\def\aap{A\&A}

\def\aaps{A\&AS}

\def\mnras{MNRAS}

\def\pasp{PASP}
\def\pasj{PASJ}

\def\nat{Nature}

\def\rmxaa{RMxAA}

\newcommand{\maihem}{\textsc{maihem}\xspace}
\newcommand{\flash}{\textsc{flash}\xspace}
\newcommand{\cloudy}{\textsc{cloudy}\xspace}

\newcommand{\ionic}[2]{#1$\,${\scshape{#2}}\xspace}
\newcommand{\ionf}[2]{#1$\,${\scshape{#2}}}

\def\arcdeg{\hbox{$^\circ$}}

\newcommand{\foiii}{[O\,{\sc iii}]\xspace}

\newcommand{\foii}{[O\,{\sc ii}]\xspace}

\newcommand{\fsii}{[S\,{\sc ii}]\xspace}

\newcommand{\fnii}{[N\,{\sc ii}]\xspace}

\newcommand{\ovi}{O\,{\sc vi}\xspace}

\newcommand{\civ}{C\,{\sc iv}\xspace}

\newcommand{\hii}{H\,{\sc ii}\xspace}

\newcommand{\heii}{He\,{\sc ii}\xspace}

\newcommand{\ha}{H$\alpha$\xspace}
\newcommand{\hb}{H$\beta$\xspace}

\definecolor{burgundy}{rgb}{0.5, 0.0, 0.13}

\usepackage{xspace}

\newcommand{\orcidicon}{\includegraphics[width=0.26cm]{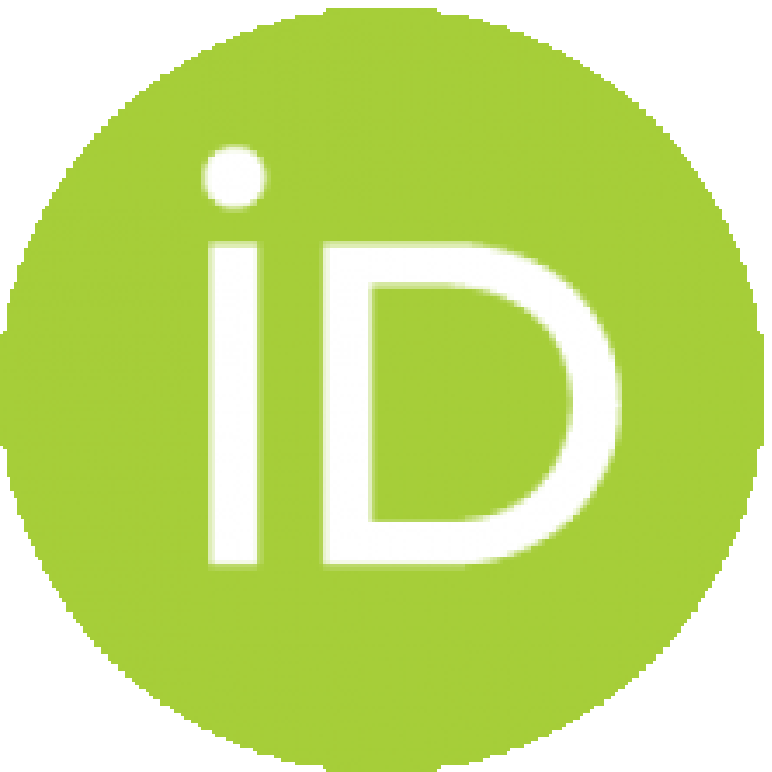}}

\newcommand{\orcidauthor}[1]{\href{https://orcid.org/#1}{\orcidicon}}

\shorttitle{Catastrophic Cooling in Superwinds. II. Exploring the Parameter Space}
\shortauthors{Danehkar et al.}

\makeatletter
\patchcmd{\frontmatter@RRAP@format}{(}{}{}{}
\patchcmd{\frontmatter@RRAP@format}{)}{}{}{}
\renewcommand\Dated@name{}
\makeatother

\begin{document}

\title{Catastrophic Cooling in Superwinds. II. Exploring the Parameter Space}

\correspondingauthor{A.~Danehkar}
\email{danehkar@umich.edu}

\author[0000-0003-4552-5997]{Ashkbiz~Danehkar}
\affiliation{Department of Astronomy, University of Michigan, 1085 S. University Ave, Ann Arbor, MI 48109, USA}

\author[0000-0002-5808-1320]{M. S. Oey}
\affiliation{Department of Astronomy, University of Michigan, 1085 S. University Ave, Ann Arbor, MI 48109, USA}

\author[0000-0001-9014-3125]{William J. Gray}
\altaffiliation{~Private address.}
\affiliation{Department of Astronomy, University of Michigan, 1085 S. University Ave, Ann Arbor, MI 48109, USA}

\date[ ]{\footnotesize\textit{Received 2021 April 14; revised 2021 July 14; accepted 2021 July 17; published 2021 November 3}}

\begin{abstract}
Superwinds and superbubbles driven by mechanical feedback from super star clusters (SSCs) are common features in many star-forming galaxies. While the adiabatic fluid model can well describe the dynamics of superwinds, several observations of starburst galaxies revealed the presence of compact regions with suppressed superwinds and strongly radiative cooling, i.e., catastrophic cooling. In the present study, we employ the non-equilibrium atomic chemistry and cooling package \maihem, built on the \flash\ hydrodynamics code, to generate a grid of models investigating the dependence of cooling modes on the metallicity, SSC outflow parameters, and ambient density. While gas metallicity plays a substantial role, catastrophic cooling is more sensitive to high mass-loading and reduced kinetic heating efficiency.  Our hydrodynamic simulations indicate that the presence of a hot superbubble does not necessarily imply an adiabatic outflow, and vice versa. Using \cloudy\ photoionization models, we predict UV and optical line emission for both adiabatic and catastrophic cooling outflows, for radiation-bounded and partially density-bounded models. Although the line ratios predicted by our radiation-bounded models agree well with observations of star-forming galaxies, they do not provide diagnostics that unambiguously distinguish the parameter space of catastrophically cooling flows.  Comparison with observations suggests the possibility of minor density bounding, non-equilibrium ionization, and/or observational bias toward the central outflow regions. 
\end{abstract}


\keywords{
\href{https://astrothesaurus.org/uat/1656}{Superbubbles (1656)};
\href{https://astrothesaurus.org/uat/1657}{Superclusters (1657)};
\href{https://astrothesaurus.org/uat/2028}{Cooling flows (2028)};
\href{https://astrothesaurus.org/uat/1565}{Star forming regions (1565)};
\href{https://astrothesaurus.org/uat/694}{H II regions (694)};
\href{https://astrothesaurus.org/uat/1570}{Starburst galaxies (1570)};
\href{https://astrothesaurus.org/uat/459}{Emission line galaxies (459)};
\href{https://astrothesaurus.org/uat/979}{Lyman-break galaxies (979)};
\href{https://astrothesaurus.org/uat/978}{Lyman-alpha galaxies (978)}
\vspace{4pt}
\newline
\textit{Supporting material:} animations, interactive figures, machine-readable tables
}


\section{Introduction}
\label{cooling:introduction}


Galactic-scale ionized gaseous outflows, the so-called ``superwinds'' as defined by \citet{Heckman1990}, are commonly observed in starburst galaxies with extreme star formation activity \citep{McCarthy1987,Heckman1987,Fabbiano1988,Heckman1990,Heckman1993,Lehnert1995,Lehnert1996,Dahlem1997a,Dahlem1997,Heckman2002,Rupke2002,Rupke2005,Martin2005,Veilleux2005}.
It has been understood that superwinds are driven by the kinetic and thermal energies deposited into the interstellar medium (ISM) by mechanical and radiative feedback from young massive stars \citep{Abbott1982,Leitherer1992,Hopkins2012} and supernova (SN) explosions \citep{Thornton1998,MacLow1999,Scannapieco2002,Creasey2013}, typically in super star clusters 
\citep[SSCs;][]{Holtzman1992,OConnell1995,Satyapal1997,Turner2003,Melo2005,Smith2006,Gilbert2007,Galliano2008}. 
These superwinds are found to be very common in intermediate- and high-redshift star-forming galaxies \citep{Pettini2001,Pettini2002,Wilman2005,Steidel2010}, so
they may play a crucial role in enriching the intergalactic medium (IGM) in the early Universe \citep{Nath1997,Lloyd-Davies2000,Ferrara2000,Heckman2002}.


The kinetic and thermal energies from OB stars displace the surrounding ISM and heat it up to $\sim 10^{7}$~K, which can result in large-scale expanding shells, the so-called ``superbubbles'' \citep{Castor1975,Bruhweiler1980,MacLow1988,Norman1989} and the production of diffuse X-ray emission \citep{Chu1990,Chu1995,Magnier1996,Strickland2002,Silich2005,Anorve-Zeferino2009}. Superbubbles have been detected around OB stellar clusters  \citep{Cash1980,Abbott1981,Cash1980,MacLow1988,Oey1995,Dove2000,Reynolds2001}, 
as well as in many star-forming galaxies \citep{Veilleux1994,Heckman1995,Marlowe1995,Weis1999,Strickland1999,Heckman2001,Sakamoto2006,Tsai2009}. 
These wind-blown bubbles carry off metal-rich material created by massive stars and SN explosions, and enrich the ISM and IGM 
with the recently produced metals. The dynamics of superbubbles have been modeled by several authors \citep[e.g.][]{Castor1975,Weaver1977,Koo1992,Koo1992a,Silich2005}.  The theoretical calculations by \citet{Weaver1977} suggested the presence of highly ionized ions such as \ionic{O}{vi} within the hot interior of the superbubble where the temperature spans the range $\sim 10^{5}$--$10^{6}$~K. Moreover, the analytical study by \citet{Silich2005} hinted at the hard (2--8\,keV) X-ray emission from the compact hot thermalized ejecta and the soft (0.3--2\,keV) diffuse X-ray emission from the superbubble interior.


The mechanical feedback from stellar winds and supernovae have been modeled using
adiabatic fluid flow by a number of authors \citep{Castor1975,Weaver1977,Chevalier1985,Canto2000}. 
The adiabatic solutions by \citet{Castor1975} and \citet{Weaver1977} were derived by taking an expanding shell surrounded by the \ionic{H}{i} and \ionic{H}{ii} regions of a uniform density that leads to the formation of a bubble. Similarly, the adiabatic steady wind solution by \citet{Chevalier1985} was obtained by adopting a continuous freely expanding stationary wind without any ambient medium.
The analytical solution of \citet[hereafter CC85;][]{Chevalier1985} describes strong superwinds driven by the kinetic and thermal energies supplied by supernovae and massive stars of a stellar cluster.
The CC85 model provided the first approximate analytic solution to the adiabatic hydrodynamic properties of the freely streaming stationary superwinds, indicating that density, temperature, and thermal pressure of the superwind around the SSC decrease with radius $r$ as $\rho_{w} \varpropto r^{-2}$, $T_{w} \varpropto r^{-4/3}$, and $P_{w} \varpropto r^{-10/3}$, respectively, while the wind velocity $u_{w}$ approaches the adiabatic terminal speed $V_{A,\infty}$. The numerical simulations of the CC85 model conducted 
by \citet{Canto2000} demonstrated that both the analytic adiabatic solutions and their numerical calculations are fully consistent for 
spherically symmetric homogeneous winds.


Moreover, the semianalytic numerical studies by \citet{Silich2003} and \citet{Silich2004} explored the impact of strongly radiative cooling, called ``catastrophic cooling'', on density, temperature, and velocity of an outflow. In particular, \citet{Silich2004} found that the wind temperature predicted by the radiative solution has a departure from the adiabatic approximation of $T_{w} \varpropto r^{-4/3}$ in massive compact clusters. 
Moreover, an increase in the ejected gas metallicity $Z$ was found to enhance radiative cooling within the cluster volume \citep{Silich2004,Tenorio-Tagle2005}. The semianalytic modeling and simulations by \citet{Tenorio-Tagle2007} estimated the locus for catastrophic cooling in the plane of cluster size ($R_{\rm sc}$) versus the mechanical luminosity ($L_{\rm mech} = \frac{1}{2} \dot{M}_{\rm sc} V_{A,\infty}^2 $), which is a function of the total stellar mass-loss rate ($\dot{M}_{\rm sc}$) and the adiabatic terminal speed ($V_{A,\infty}$).

Recent observations are reported to exhibit signs of catastrophic cooling and the
suppression of superwinds and superbubbles in SSCs in several
starburst galaxies: M82 \citep{Smith2006,Westmoquette2014}, NGC 2366
\citep{Oey2017}, NGC 5253 \citep{Turner2017}, where the SSCs
are embedded within 1--2 pc, dense, ultracompact, high-pressure gas.
Similarly, the the most extreme Green Peas
\citep[GPs;][]{Jaskot2017} show kinematics consistent with suppressed
superwinds, contrary to other starbursts.
Recent ALMA observations of distant star-forming galaxies reveal
the presence of extended [\ionf{C}{ii}] halos that could also be
produced by starburst-driven cooling outflows 
\citep{Fujimoto2019,Fujimoto2020,Pizzati2020}.  
These starburst-driven strongly cooled outflows could be explained by the radiative solution semianalytically derived by \citet{Silich2003,Silich2004}. 
More recently, \citet{Gray2019a} investigated the occurrence of catastrophic cooling through hydrodynamic simulations including radiative non-equilibrium cooling, which showed the strong enhancement of highly ionized ions such as \ionic{C}{iv} and \ionic{O}{vi}. 


In the recent work by \citet[][Paper I]{Gray2019a}, the hydrodynamic simulations with the non-equilibrium atomic chemistry and cooling package \maihem \citep{Gray2015,Gray2016,Gray2019} were performed assuming a default solar value for the metallicity $Z$ of the outflow, and a typical mass-loss rate of $10^{-2}$ M$_{\odot}$\,yr$^{-1}$. As found by \citet{Silich2004}, \citet{Tenorio-Tagle2005}, and \citet{Gray2019}, the metallicity can have a significant role in strongly enhancing radiative cooling. Moreover, the mechanical luminosity, i.e. the mass-loss rate and wind terminal speed can affect the domain of catastrophic cooling \citep{Tenorio-Tagle2007}.

In this paper, we therefore investigate a larger parameter space to explore the
effect of metallicity $Z$, mass-loss rate $\dot{M}_{\rm sc}$, and wind
terminal velocity $V_{\infty}$  on the behavior of
catastrophic cooling and superwind suppression. Although our models are
parameterized in terms of $\dot{M}_{\rm sc}$ and $V_{\infty}$, we note
that these are the effective values that may result after mass-loading
or heating efficiency effects.  As in Paper~I, we use one-dimensional, spherically
symmetric hydrodynamic simulations coupled to radiative thermal
functions using the non-equilibrium atomic chemistry and cooling
package \maihem\ \citep{Gray2019a} to run an extensive grid of models.
In addition, we also investigate some emission-line
diagnostics for these conditions.  These are generated by
post-processing the \maihem\ models with the \cloudy\ photoionization code.

This paper is organized as follows.  Section \ref{cooling:winds} describes the configuration of our hydrodynamic simulations of superwinds, including the radiative thermal functions, initial and boundary conditions. Section \ref{cooling:photoionization} describes the implementation of the photoionization modeling, as well as computing population synthesis models for SSCs. The hydrodynamic results from our \maihem simulations are presented in Section~\ref{cooling:hydrodynamics:results}, followed by the emission lines calculated by \cloudy and diagnostic diagrams in Section~\ref{cooling:cloudy:results}. Our results  are compared with observations in Section~\ref{cooling:observations}, and are discussed in Section~\ref{cooling:discussion}. Our conclusions are given in Section~\ref{cooling:conclusion}. 

\section{Hydrodynamic Simulations}
\label{cooling:winds}

We solve the hydrodynamic equations coupled to the radiative cooling and heating functions using the non-equilibrium atomic chemistry and cooling package \maihem \citep[Models of Agitated and Illuminated Hindering and Emitting Media;][]{Gray2015,Gray2016,Gray2019}, which is a modified version of the adaptive mesh hydrodynamics code \flash v4.5 \citep{Fryxell2000}. We use the directionally unsplit pure hydrodynamic solver \citep{Lee2009,Lee2009a,Lee2013} 
together with the second-order Monotone Upwind-centered Scheme for Conservation Laws (MUSCL)--Hancock scheme \citep{vanLeer1979} developed for arbitrary Lagrangian Eulerian (ALE) and smoothed particle hydrodynamics (SPH) methods.\footnote{\flash solves fluid equations using either (1) a directionally unsplit pure hydrodynamic solver, or (2) a directionally split piecewise-parabolic method (PPM) solver; together with one of reconstruction schemes, namely first-order Godunov, second-order MUSCL--Hancock, third-order PPM, and fifth-order Weighted Essentially Non-Oscillatory (WENO).} 
To avoid odd--even instabilities, we also employ a hybrid type of the Riemann solver \citep{Toro1994}, combining the Roe solver \citep{Roe1981} and the Harten--Lax--van Leer--Einfeldt (HLLE) solver \citep{Einfeldt1988,Einfeldt1991}. The HLLE solver is a modified version of the HLL solver that is positively conservative under certain wavespeed conditions in order to provide more stable and diffusive solutions in strongly shocked regions. 

Following \citet{Silich2004}, the equations of hydrodynamics applicable to spherically symmetric superwinds in a steady
state with the radiative cooling for a spherically symmetric SSC of radius $R_{\rm sc}$ without accounting for the gravitational attraction from the SSC are:
\begin{equation}
\frac{1}{r^2} \frac{d}{dr} \left( \rho_{w} u_{w} r^2 \right) = q_{m}, \label{eq:1}%
\end{equation}
\begin{equation}
\rho_{w} u_{w} \frac{d u_{w}}{dr} + \frac{d P_{w}}{dr} = -q_{m}u_{w}, \label{eq:2}%
\end{equation}
\begin{equation}
\frac{1}{r^2} \frac{d}{dr} \left[ \rho_{w} u_{w} r^2  \left( \frac{u^{2}_{w}}{2} +\frac{\gamma}{\gamma -1} \frac{P_{w}}{\rho_{w}}  \right) \right]= q_{e} - q_{c} + q_{h}, \label{eq:3}%
\end{equation}
where $r$ is the radial distance,
$u_w$ the wind velocity, $\rho_w$ the wind density, $P_w$ the wind thermal pressure, 
$q_{m} = 3 \dot{M}_{\rm sc} / 4 \pi R^3_{\rm sc}$ the total mass-loss rate per unit volume, $q_{e} = 3 \dot{E}_{\rm sc} / 4 \pi R^3_{\rm sc}$ the total energy deposition rate per unit volume, $\dot{M}_{\rm sc}$ the total mass-loss rate, $\dot{E}_{\rm sc}$ the total energy deposition rate, 
$\gamma=5/3$ the ratio of specific heats, 
$q_{c} =n_w^2 \Lambda (Z,T_{w})$ the cooling rate per unit volume, $q_{h} =n_w \Gamma (Z)$ the heating rate per unit volume, 
$n_w$ the wind number density, $\Lambda (Z,T_{w})$ the cooling function \citep[][]{Raymond1976,Oppenheimer2013},
$\Gamma(Z)$ the heating function \citep[see e.g.][]{Wolfire2003,Gnedin2012,Gray2016,Gray2019}, 
$T_{w}$ the wind temperature, and $Z$ the metallicity. 
We adopt the gamma-law equation of state for an ideal gas, $P_{w}=
(\gamma -1 ) \rho_{w} \epsilon_{w}$, where $\epsilon_{w}$ is the
internal energy.  This accounts for changes in the mean atomic mass
due to evolution of the ionization states of all the species.
Outside the SSC ($r > R_{\rm sc}$), $q_{m}$ and $q_{e}$ vanish
(CC85).  The wind model assumes that radiative thermal effects
within the SSC are negligible, so $q_{c}$ and $q_{h}$ vanish at $r <
R_{\rm sc}$. Taking $q_{c}=0$ and $q_{h}=0$ at $r > R_{\rm sc}$, we
recover the adiabatic solution in CC85 and \citet{Canto2000}. The wind
terminal velocity is also defined as $V_{\infty} = ( 2 \dot{E}_{\rm
  sc} / \dot{M}_{\rm sc} )^{1/2}$. 

Assuming the wind of a uniform density, solving Eqs. (\ref{eq:1})--(\ref{eq:3}) yields the wind density and pressure at $r>R_{\rm sc}$ away from the SSC \citep{Silich2004}:
\begin{equation}
\rho_{w} = \frac{\dot{M}_{\rm sc}}{ 4 \pi u_{w} r^2} , \label{eq:4}%
\end{equation}
\begin{equation}
\frac{d P_{w}}{dr} = \frac{\dot{M}_{\rm sc}}{4 \pi r^3} \frac{(\gamma -1) r (q_{h}-q_{c}) - 2 \gamma u_w P_w}{\rho_w (u_w^2 -c_s^2) }, \label{eq:5}%
\end{equation}
where $c_s=(\gamma P_{w} / \rho_{w})^{1/2} =(\gamma k_{\rm B} T_{w} / \mu ) ^{1/2}$ is the sound speed, $\mu$ the mean mass per wind particle,  and $k_{\rm B}$ the Boltzmann constant.  
Following the stationary wind solution of \citet{Silich2004}, at $r=R_{sc}$, the Mach number $M = u_w / c_s =1$ (CC85), so the wind velocity $u_w$ is equal to the sound speed $c_s$. Following CC85, the wind velocity is $u_w = V_{\infty} / 2$ at the cluster boundary $r=R_{sc}$. Eq. (\ref{eq:4}) implies that the wind density at the cluster boundary should be $\rho_{w} = \dot{M}_{\rm sc}/ 2 \pi V_{\infty} R_{\rm sc}^2$. As $c_s=u_w$ at $r=R_{sc}$, the wind temperature and pressure at the cluster boundary should be $T_w =  \mu  V_{\infty}^2 / 4 \gamma k_{\rm B} $ and $P_w = \rho_w  V_{\infty}^2 / 4 \gamma$, respectively. These values of the wind density and temperature are used as the boundary conditions at $r=R_{sc}$ in our hydrodynamic simulations as described in the following subsection. 

\subsection{Initial and Boundary Conditions}
\label{cooling:configurations}

The hydrodynamic simulations are computed from $t=0$ until an age of 1
Myr at checkpoint intervals of 0.1 Myr.  This final age is long enough
to generate a mature outflow, but too short 
for any appreciable evolution in the SED and stellar population.  It
describes typical superwinds driven by extremely young starbursts.
The hydrodynamic model employs one-dimensional, spherical geometry with the initial radius equal to the SSC $R_{\rm sc}$, which is set to 1 pc in our study.
This corresponds to the flow injection radius. The outer boundary is set to 250 pc in our hydrodynamic simulations that covers the expansion of fast winds with high mass-loss rates at their final age. 
We note that the associated Str\"{o}mgren spheres are often larger, and the outer boundary in our \cloudy\ models are unlimited in radius (described in \S\,\ref{cooling:photoionization}). The hydrodynamic models are simulated on a base grid consisting of $N_{x}=512$ blocks that are allowed up to two levels of adaptive mesh refinement. This gives a maximum resolution of 0.244\,pc. 
The adaptive mesh algorithm \citep[\textsc{paramesh};][]{MacNeice2000} employed by \flash uniformly covers the physical space, whose refinement criterion is an error estimator \citep{Loehner1987} that is configured to follow the changes in density, temperature, pressure, and velocity. The ambient medium surrounding the SSC does not have any initial kinematics ($u_w=0$).

The initial conditions of the wind velocity, temperature, and density at the cluster boundary for the hydrodynamic simulations are adopted based on the analytic radiative solutions \citep{Silich2004}, which are similar to the CC85 adiabatic solutions. The initial values of the temperature and density in the surrounding regions are set to the temperature $T_{\rm amb}$ -- that is determined by photoionization of the SSC -- and density $n_{\rm amb}$ of the ambient medium, respectively. The initial wind density $\rho_{w, {\rm sc}}$ and temperature $T_{w, {\rm sc}}$ at the cluster boundary ($r=R_{\rm sc}$)  are then calculated by assuming the analytic solution:
\begin{equation}
\rho_{w, {\rm sc}} = \frac{ \dot{M}_{\rm sc}}{ 4 \pi  R_{\rm sc}^2 u_{w, {\rm sc}}} ,
\label{eq:8}%
\end{equation}
\begin{equation}
T_{w, {\rm sc}} =  \frac{ 1 }{\gamma}  \frac{\mu }{ k_{\rm B}} u_{w, {\rm sc}}^2 ,
\label{eq:9}%
\end{equation}
where $u_{w, {\rm sc}}$ is the wind velocity at $r=R_{\rm sc}$ given by the wind terminal speed $V_{\infty}$  \citep{Chevalier1985,Canto2000}:
\begin{equation}
u_{w, {\rm sc}} = \tfrac{1}{2} V_{\infty} . 
\end{equation}
We initialize the boundary wind velocity to $V_{\infty} / 2$. This initial wind velocity configuration is based on the wind terminal speed that easily allows us to compare the radiative solution to the adiabatic solution. 
For the effective area of the outflow $\Omega R^2_{\rm sc}$ in Eq. (\ref{eq:8}), we use $\Omega = 4 \pi$ corresponding to a fully isotropic outflow, 
while \citet{Gray2019a} adopted $\Omega = \pi$ describing an anisotropic broken outflow extended perpendicularly to the galactic plane. 

Following the CC85 adiabatic solutions, the wind velocity at $r=R_{\rm sc}$ is assumed to be equal to the local sound speed, $u_{w, {\rm sc}} = c_{s,\rm sc}$, which corresponds to a Mach number of 1. The wind terminal speed $V_{\infty}$  is related to the mechanical luminosity as 
\begin{equation}
L_{\rm mech} = \tfrac{1}{2} \dot{M}_{\rm sc} V_{\infty}^2 . 
\end{equation}
After setting the initial temperature and density from the assumed
ambient density, the thermal pressure and internal energy are determined
by solving equations
(\ref{eq:1})--(\ref{eq:3}) in \maihem, using the equation of state.
Thus, the wind solution develops from these initial conditions.

The initial thermal pressure derived from the hydrodynamic simulation is
found to be consistent with the analytically derived initial pressure: 
\begin{equation}
P_{w, {\rm sc}} = \frac{ 1 }{\gamma} \frac{ \dot{M}_{\rm sc} u_{w, {\rm sc}}}{ 4 \pi  R_{\rm sc}^2 }  .
\label{eq:10}%
\end{equation}
Alternatively, we could employ the initial analytic pressure $P_{w, {\rm sc}}$ and density $\rho_{w, {\rm sc}}$ and determine the initial temperature $T_{w, {\rm sc}}$ by solving the equation of state in \maihem that would provide the same analytic temperature given in Eq. (\ref{eq:9}). The internal energy per unit mass can also be calculated from the internal pressure and density as follows:
\begin{equation}
\epsilon_{w, {\rm sc}} = \frac{1}{(\gamma-1)} \frac{P_{w, {\rm sc}}  }{ \rho_{w, {\rm sc}} }  ,
\label{eq:11}%
\end{equation}
which is calculated by solving the equation of state in \maihem. The internal and kinetic energies per unit mass are used to initialize the total energy per unit mass at the cluster boundary ($E_{w, {\rm sc}}= \epsilon_{w, {\rm sc}} + \frac{1}{2} |u_{w, {\rm sc}} |^2$).

Following \citet{Gray2019a,Gray2019}, the gas ejected by the cluster
at $R_{\rm sc}$ is approximated to be in collisional ionization
equilibrium (CIE) at the boundary, so it depends only on the initial
wind temperature and ionizing UV background. The photoionization code
\cloudy is called to generate the CIE ionization fractions that are a
function of temperature and applicable ionizing UV background. 
Our procedure makes tabulated grids of ionization fractions for all the ions in CIE that span the temperature range of $10^2$--$10^8$\,K, which are linearly interpolated over to yield the ionization fractions at the outflow boundary for a given temperature during our hydrodynamic simulation.
The tabulated grids incorporate all the processes considered by \cloudy, including collisional excitations,
recombination, and collisional ionization. 
The ionization fractions of the ejected gas at the outflow boundary are then calculated by \maihem at runtime through an interpolation on the CIE tabulated grids generated by \cloudy.

Table~\ref{tab:sim:param} summarizes the parameters used in our \maihem hydrodynamic simulations that includes the metallicity used in \maihem simulations ($Z$), the mass-loss rate ($\dot{M}_{\rm sc}$), and the wind terminal speed ($V_{\infty}$) in the first three columns. It also lists the metallicities used in Starburst99 models for the Geneva-Rot stellar evolution, high-resolution spectra, and UV spectrum (columns 4--6). The last three columns present Starburst99 outputs: the total luminosity ($L_{\rm tot}$), the fraction of ionizing photons in the \ionic{H}{i} continuum relative to the total luminosity, $f($H$^{+})$, and the ionizing luminosity $L_{\rm ion}$.
We consider the gas metallicity of $\hat{Z} \equiv Z/$Z$_{\odot}=1$, $0.5$, $0.25$, and $0.125$, where $Z/$Z$_{\odot}=1$ corresponds to the ISM abundances listed in Table~\ref{tab:abund:deplet}. 
We adopt the typical ISM abundances of \citet{Savage1977}, together
with the ISM gas-phase oxygen abundance of \citet{Meyer1998}, as the
baseline solar metallicity (see Table~\ref{tab:abund:deplet}).  
Typical ISM grains with a dust-to-metal mass ratio of $M_{\rm d}/M_{\rm Z}=0.2$ are
incorporated into our initial models of the ionization fraction computed by \cloudy\
(see also \S\,\ref{cooling:photoionization}).  Dust
grains are not included as a distinct species in the
\maihem\ simulations, but we use C/O ratios that account for depletion
as described in \S\,\ref{cooling:photoionization}.
For the fiducial models with the solar metallicity ($Z/$Z$_{\odot}=1$), we calculate the boundary density $\rho_{w, {\rm sc}}$ and temperature $T_{w, {\rm sc}}$ using the total mass-loss rate $\dot{M}_{\rm sc}$ of $10^{-1}$, $10^{-2}$, $10^{-3}$, and $10^{-4}$\,M$_{\odot}$\,yr$^{-1}$, and the wind terminal speed $V_{\infty}$ of 250, 500, and 1000\,km\,s$^{-1}$. 
For the models with sub-solar metallicities, namely $Z/$Z$_{\odot}=0.5$, $0.25$, and $0.125$, we scale the mass-loss rates and wind velocities of the solar models according to $ \dot{M}_{\rm sc} \varpropto Z^{0.72}$ \citep{Mokiem2007} and $V_{\infty} \varpropto Z^{0.13}$ \citep{Vink2001}, since we are interested only in extremely young, pre-supernova SSCs that are dominated by stellar winds.

For the ambient medium in our \maihem\ simulations, we adopt the
photoionized temperature structure created by the ionizing SED from the
parent SSC.  This is done using a single \cloudy\ model applied to the
density profile predicted by \maihem\ from a preliminary simulation,
for the fiducial age of 1 Myr, described in detail \S\,\ref{cooling:starburst99}
(the ``PI model'').  This sets the ionized radius for
the final \maihem\ model, and determines how long the outflow expands
into an ionized medium.  For most models, the ambient medium remains
ionized at 1 Myr.
For optically thick models transitioning to a neutral ISM (see\,\S\ref{cooling:cloudy:results}), which are mostly those with
$n_{\rm amb}=10^3\ \rm cm^{-3}$, we adopt an ambient
temperature of $T_{\rm amb} = 5 \times 10^3$\,K.  
The ionizing SED is generated
by the Starburst99 population synthesis model described in 
\S~\ref{cooling:photoionization} for the fiducial model of $M_\star =
2.05\times10^6$\,M$_{\odot}$ at age 1\,Myr.  This SED is also
included as a background UV radiation field in the \maihem\ simulation
by using a simple $r^{-2}$ decrease in flux.  \maihem\ does not
calculate detailed radiative transfer, but since the wind is low density
and therefore optically thin, this is a reasonable assumption.
Initially, the entire surrounding domain outside the SSC is assumed to
have ambient densities of $n_{\rm amb}=1$, $10$, $10^2$, and
$10^3$\,cm$^{-3}$.

\begin{table*}
\begin{center}
  \caption[]{Summary of Parameters in Hydrodynamic Simulations by \maihem and Stellar Population Models by Starburst99.
    \label{tab:sim:param}
    }
\footnotesize
\begin{tabular}{c|rc|ccc|ccc}
  \hline\hline\noalign{\smallskip}
\multicolumn{1}{c|}{Hydro}&\multicolumn{2}{c|}{\maihem Outflow Parameters}&\multicolumn{3}{c|}{Starburst99 Inputs}&\multicolumn{3}{c}{Starburst99 Output}\\
\noalign{\smallskip}
$Z$	&\multicolumn{1}{c}{$\dot{M}_{\rm sc} (\varpropto Z^{0.72})$}& \multicolumn{1}{c|}{$V_{\infty} (\varpropto Z^{0.13})$}	&$Z$	&$Z$	&$Z$ &$\log L_{\rm tot}$	& $f($H$^{+})$	&$\log L_{\rm ion}$ \\
(Z$_{\odot}$)	&\multicolumn{1}{c}{(M$_{\odot}$ yr$^{-1}$)}&\multicolumn{1}{c|}{(km s$^{-1}$)}	& Geneva-Rot	&High Res.Spec.	&UV Spec. &(erg s$^{-1}$)	&	&(erg s$^{-1}$)	\\
\noalign{\smallskip}
\tableline
\noalign{\smallskip}
1.0	&1.0 $\times$ [$10^{-1}$, $10^{-2}$, $10^{-3}$, $10^{-4}$]	&[250, 500, 1000] &0.014	&0.020	&Solar	&42.97	&0.41	&42.58	\\
\noalign{\smallskip}
0.5	&0.607 $\times$ [$10^{-1}$, $10^{-2}$, $10^{-3}$, $10^{-4}$]	&[229, 457, 914] &0.008	&0.008	&LMC/SMC	&42.97	&0.42	&42.60	\\
\noalign{\smallskip}
0.25	&0.369 $\times$ [$10^{-1}$, $10^{-2}$, $10^{-3}$, $10^{-4}$]	&[209, 418, 835] &0.008	&0.008	&LMC/SMC	&42.97	&0.42	&42.60\\
\noalign{\smallskip}
0.125	&0.224 $\times$ [$10^{-1}$, $10^{-2}$, $10^{-3}$, $10^{-4}$]	&[191, 382, 736] &0.002	&0.008	&LMC/SMC	&42.98	&0.50	&42.67\\
\noalign{\smallskip}\hline
\end{tabular}
\end{center}
\begin{tablenotes}
\footnotesize
\item[1]\textbf{Note.} The ambient temperature ($T_{\rm amb}$) is set to the value calculated by the \textsc{cloudy} model
for the given Starburst99 SED. 
Other \maihem parameters are as follows: the cluster radius $R_{\rm sc}=1$\, pc, the ambient density $n_{\rm amb}=1$, $10$, \ldots, $10^{3}$\,cm$^{-3}$, the number of blocks $N_{x}=512$, and time intervals $t=0.1$, 0.2, \ldots, $1.0$\,Myr. 
Other Starburst99 parameters are as follows: the total stellar mass $M_{\star}= 2.05 \times 10^6$\,M$_{\odot}$,
and age $t=1$\,Myr. 
\end{tablenotes}
\end{table*}

\subsection{Non-equilibrium Cooling Functions}
\label{cooling:winds:radiative:nei}

To include hydrodynamic heating and background UV radiation, the
cooling and heating functions of atomic species are calculated using
non-equilibrium ionization (NEI) conditions. The cooling rate per volume
$q_{c}(Z,T_{w})$ is determined by the \maihem\
cooling routine using the cooling function $\Lambda (Z,T_{w})$
based on the gas metallicity $Z$, in addition to the gas temperature
$T_{w}$ found by hydrodynamic solutions for given physical conditions
and background radiation.  
We use the cooling routine implemented by \citet{Gray2015}, which extended
the ion-by-ion cooling efficiencies $\Lambda_i (T_w)$ of \citet[][]{Gnat2012} down to 5000\,K, for the given ions. 
Additionally, \citet{Gray2015} included NEI in \maihem, which can track non-equilibrium ionization and recombination of species. 
Furthermore, as fully described by \citet{Gray2019}, the \maihem package takes account of the heating rate per volume, $q_{h}(Z)$, by calculating the heating function $\Gamma (Z)$ for the outflow of metallicity $Z$. This atomic chemistry was implemented by incorporating collisional ionization rates \citep{Voronov1997}, radiative recombination rates \citep{Badnell2006}, and dielectronic recombination rates \citep[see Table 1 in][]{Gray2015}. 
With the inclusion of the atomic chemistry package, our simulations are performed using the multispecies extension for the equation of state that accounts for the change in average atomic weight on the thermodynamic quantities. The species and non-equilibrium cooling were incorporated into \maihem. These two capabilities, namely non-equilibrium ionization and radiative thermal function allow us to use \maihem to investigate the radiative solution of the superwind model.

The package \maihem computes the photoionization and photoheating rate for each ion \citep{Verner1995,Verner1996} to include the effect of an ionizing UV background \citep{Gray2016}. 
The most recent version of \maihem updated by \citet{Gray2019} is able to track 
non-equilibrium ionization and recombination of 84 species across 13 elements, namely 
hydrogen (\ionic{H}{i}--\ionic{H}{ii}), helium (\ionic{He}{i}--\ionic{He}{iii}), carbon (\ionic{C}{i}--\ionic{C}{vii}), 
nitrogen (\ionic{N}{i}--\ionic{N}{viii}), oxygen (\ionic{O}{i}--\ionic{O}{ix}), neon (\ionic{Ne}{i}--\ionic{Ne}{xi}), 
sodium  (\ionic{Na}{i}--\ionic{Na}{vi}), magnesium (\ionic{Mg}{i}--\ionic{Mg}{vi}), 
silicon (\ionic{Si}{i}--\ionic{Si}{vi}), sulfur (\ionic{S}{i}--\ionic{S}{vi}),
argon (\ionic{Ar}{i}--\ionic{Ar}{vi}), calcium (\ionic{Ca}{i}--\ionic{Ca}{vi}), and 
iron (\ionic{Fe}{i}--\ionic{Fe}{vi}). 
The cooling efficiencies of the new ions in the expanded network of \citet{Gray2019} were computed 
using \cloudy, in addition to the method used by \citet{Gray2015}. 
The tabulated grids of cooling efficiencies are read into \maihem\ 
during the initialization for each model.
The latest version models include the column densities of \ionic{N}{v}
and \ionic{O}{vi} \citep{Gray2019}.

As mentioned in \citet{Gray2019a}, the cooling efficiency $\Lambda_i (T_w)$ of species $i$, the number density $n_i$ of species $i$, and the number of electrons $n_e$ yield the cooling rate per volume:
\begin{equation}
q_{c}(Z, T_w) = n_w^2\Lambda(Z, T_w)  = \sum_{i}^{}  n_i n_e \Lambda_i (T_w),
\label{eq:7}%
\end{equation}
which is used in the computations of fluid models to provide radiative cooling. As the ion-by-ion cooling efficiencies $\Lambda_i (T_w)$ are calculated based on the temperature determined by the hydrodynamic simulations for specified physical conditions and background radiation, we see how the mechanical feedback (via $\dot{M}_{\rm sc}$ and $V_{\infty}$) can lead to the catastrophic cooling regime. Furthermore, the cooling function $\Lambda(Z, T_w)$ is a function of the metallicity specified for the ejected gas. 

The heating rate per volume is calculated using the photo-heating efficiency $\Gamma_i$  and number density $n_i$ of species $i$ as follows \citep[][]{Gray2019}:
\begin{equation}
q_{h}(Z) = n_w \Gamma(Z) = \sum_{i}^{}  n_i  \Gamma_i,
\label{eq:7:2}%
\end{equation}
where the ion-by-ion photo-heating efficiencies $\Gamma_i$ are estimated by \maihem \citep[see][]{Gray2016} using a background UV spectral energy distribution (SED) and the photoionization cross section of species $i$, taken from \citet{Verner1995} and \citet{Verner1996}. 

\begin{figure*}
\begin{center}
\includegraphics[width=0.6\textwidth, trim = 0 0 0 0, clip, angle=270]{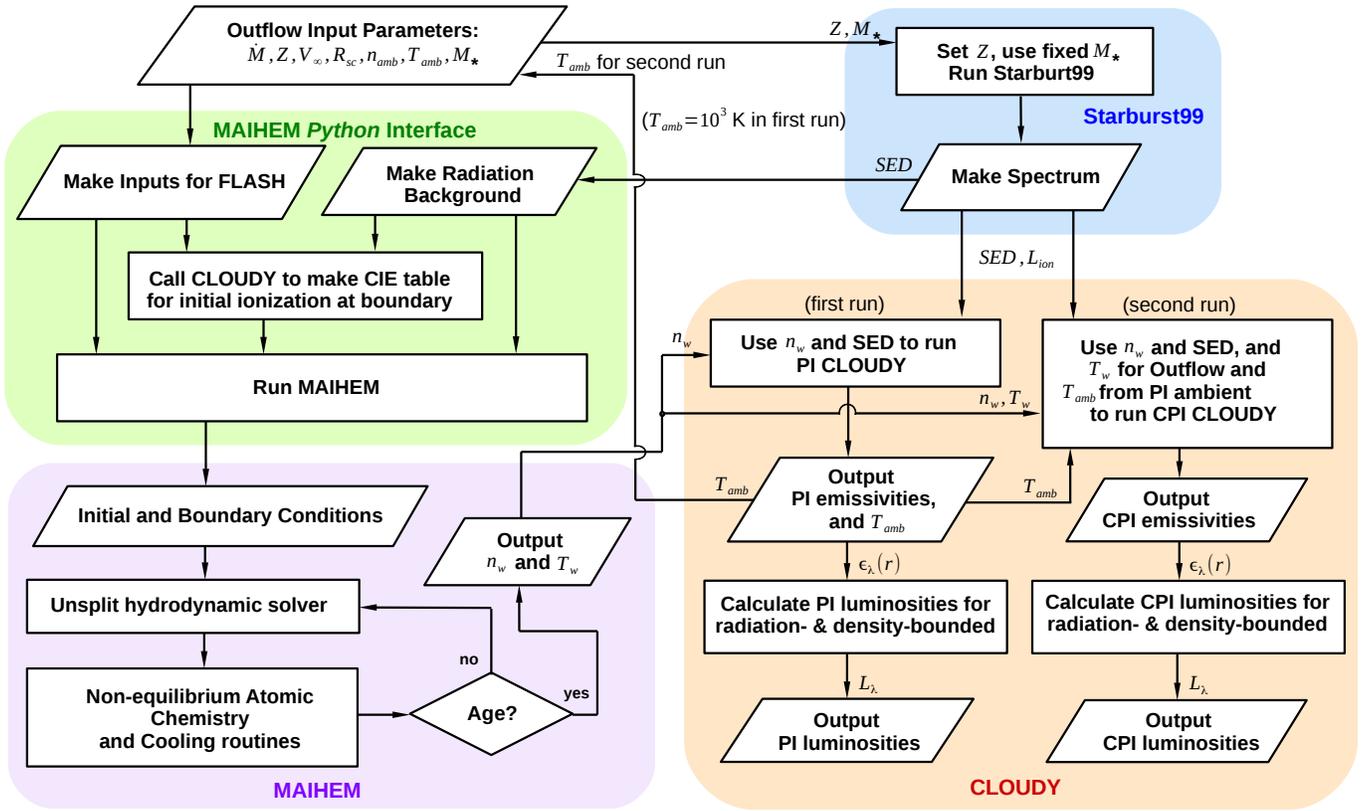}
\end{center}
\caption{Flowchart of the \textsc{maihem}-\textsc{cloudy} interface
implemented in our hydrodynamic simulations. 
Starburst99 produces the radiation spectrum for the total stellar mass ($M_{\star}= 2.05 \times 10^6$\,M$_{\odot}$) and given metallicity ($Z$), which is used 
by \textsc{maihem} and \textsc{cloudy}. The \maihem python interface
calls \cloudy to generate the CIE tables as a function of
temperature for the cluster boundary, and produce the background UV SED 
from the Starburst99 synthetic stellar spectrum to be used by NEI atomic chemistry 
and cooling routines in \maihem. Solving the hydrodynamic fluid equations, together with 
radiative cooling and heating functions, \maihem produces the outflow
temperature ($T_{w}$), density profile ($n_{w}$), and NEI
states according to the Starburst99 UV SED. 
The PI \cloudy model (pure photoionization) uses only the density profile made by the \textit{first} \maihem hydrodynamic run with 
an isothermal ambient temperature of $10^3$\,K, together with 
the SED model produced by Starburst99. 
The ambient temperature profile ($T_{\rm amb}$) predicted by the PI \cloudy model is employed for the ambient medium in the \textit{second} \maihem hydrodynamic run and the CPI case.
The CPI \cloudy model (photoionization + hydrodynamic collisional ionization)
 is calculated in the same way, but uses both
the temperature and density profiles of the outflow model produced by the second \maihem run, and   
the ambient temperature structure from the PI model. 
The radiation- and density-bounded luminosities
$L_{\lambda}$ are calculated from the emissivities $\epsilon_{\lambda} (r)$ produced by the PI amd CPI \cloudy models, described in Appendix~\ref{appendix:a}.
\label{fig:maihem:flowchart}
}
\end{figure*}

\section{Photoionization Calculations}
\label{cooling:photoionization}

We next carry out photoionization modeling to calculate emissivities of a set of UV/optical
emission lines that are typically used for diagnostic purposes. These
are used to build diagnostic
diagrams (\S\,\ref{cooling:opt:diagnostics}), which we compare with
observations (\S\,\ref{cooling:observations}). 
The photoionization modeling is implemented as post-processing using the program \cloudy v17.02, which is described by  
\citet{Ferland1998,Ferland2013,Ferland2017}. The outflow temperature $T_{w}$ and density $n_{w}$ as a function of the radius $r$ obtained from our hydrodynamic simulations are used as inputs to our photoionization models. Additionally, the SED of the SSC generated by Starburst99 \citep{Leitherer1999,Leitherer2014}, together with its predicted ionizing luminosity, as described in \S\,\ref{cooling:starburst99}, is used as an ionizing 
source for our photoionization models. 
The stellar evolution and atmosphere models employed by Starburst99 use different increments in metallicity, and so the metallicity in each of our photoionization models is set to the closest match with the metallicity $Z$ (i.e., $Z/$Z$_{\odot}=1$, $0.5$, $0.25$, and $0.125$) used in the corresponding hydrodynamic simulation, as shown in Table~\ref{tab:sim:param}. 

Similar to \citet{Gray2019a}, we model the ionization states for two
cases using the same hydrodynamically generated density distribution
from \maihem:
a purely photoionized model based on resetting the entire gas
distribution to neutral temperature, and therefore without any hydrodynamic thermal
contributions (PI case); and a model with both photoionization and
hydrodynamic collisional ionization (CPI case), based on the
\maihem\ temperature distribution. These are analogous to
models B and C in Paper~I.  We use a single Cloudy model to
generate a more realistic ionization structure, although it is necessarily CIE, since
Cloudy cannot model non-CIE. 
We summarize these models as follows (Figure~\ref{fig:maihem:flowchart}): 
\begin{enumerate}
\item PI case:
  This model is only used to generate inputs for the CPI case.
The ambient temperature in \maihem is initially set to an
  isothermal value of 1000\,K to estimate initial density
  profiles.  Together with the ionizing SED and luminosity from
  Starburst99, they are
 employed by \cloudy\ to calculate the photoionization and
  emissivities due to the stellar radiation of the SSC.  
  The \cloudy\ calculations stop when the temperature drops below 950\,K.
    The resulting temperature profile in the PI case is then used to
  obtain the photoionized radius and mean ambient temperature 
  for the ionized portion of the ambient medium and 
  shell in the CPI case (see below, and Section~\ref{cooling:configurations}). 

\item CPI case: Both the density and temperature profiles
  generated by \maihem together with the ionizing SED and luminosity
  produced by Starburst99 are used by \cloudy to compute the
  ionization states and emissivities resulting from both the stellar
  radiation of the SSC and thermal contributions from the hydrodynamic
  outflow (see Figure~\ref{fig:maihem:flowchart}).
  The temperature profile found in the PI case is applied to the
ambient medium and defines the radius at which the model transitions
to neutral and becomes optically thick.  The mean ambient ionized
temperature in the PI case is adopted for running the
\maihem\ model.  It is also adopted for the temperature in the ambient
medium surrounding the shell, as well as the ionized, isothermal
region of the shell, roughly $\sim 1$--2\,pc outward from the shell
inner boundary.  The CPI calculations stop according to the optical
depth (H$^{+}$) defined by the PI model.
\end{enumerate}

Thus, the PI \cloudy\ model is used to define the ambient temperature
profile in the ambient medium for the
hydrodynamic simulation by \maihem (see \S\,\ref{cooling:winds}), as well as 
the ambient medium in the CPI \cloudy model. 
The \maihem simulation produces the non-equilibrium ionization states and temperature profile of the outflow. 
However, we only employ the temperature profile generated by \maihem
for the CPI model, and not the ionization states, so our \cloudy\ models calculate only
CIE conditions.  NEI conditions, which can be estimated by including both the 
ionization states and temperature structure predicted by \maihem,
generate higher fluxes in certain highly-ionized lines such as
\ionic{C}{iv} and \ionic{O}{vi}, as shown in Paper~I. 
Our CPI models, which combine the thermal effects
and density distribution from both the hydrodynamic outflow and
photoionization, are the final, realistic models.

\begin{table}
\begin{center}
\caption[]{ISM abundances and depletion factors for \cloudy models
\label{tab:abund:deplet}}
\footnotesize
\begin{tabular}{lcc}
  \hline\hline\noalign{\smallskip}
Element	& $n({\rm X})/n({\rm H})$ & $(1 - f_{\rm dpl}({\rm X}))$ \\
\noalign{\smallskip}
\tableline
\noalign{\smallskip}
He   &   $0.098$  &  1.0   \\
\noalign{\smallskip}
C    &  $2.87 \times 10^{-4}$   &  0.4   \\
\noalign{\smallskip}
N    &  $7.94 \times 10^{-5}$   &  1.0   \\
\noalign{\smallskip}
O    &  $3.19 \times 10^{-4}$   &  0.6   \\
\noalign{\smallskip}
Ne   &  $1.23 \times 10^{-4}$   &  1.0   \\
\noalign{\smallskip}
Na   &  $3.16 \times 10^{-7}$   &  0.2   \\
\noalign{\smallskip}
Mg   &  $1.26 \times 10^{-5}$   &  0.2   \\
\noalign{\smallskip}
Si   &  $3.16 \times 10^{-6}$   &  0.03  \\
\noalign{\smallskip}
S    &  $3.24 \times 10^{-5}$   &  1.0   \\
\noalign{\smallskip}
Ar   &  $2.82 \times 10^{-6}$   &  1.0   \\
\noalign{\smallskip}
Ca   &  $4.10 \times 10^{-10}$   &  $10^{-4}$   \\
\noalign{\smallskip}
Fe   &  $6.31 \times 10^{-7}$   &  $10^{-2}$   \\
\noalign{\smallskip}\hline
\end{tabular}
\end{center}
\begin{tablenotes}
\footnotesize
\item[1]\textbf{Note.} The abundance $n({\rm X})/n({\rm H})$ of the element ${\rm X}$ is by number relative to the hydrogen abundance. The depletion factor $f_{\rm dpl}({\rm X})$ specifies the depletion fraction of the element ${\rm X}$ onto dust grains. The carbon abundance is set in a way to provide $[n({\rm C})(1 - f_{\rm dpl}({\rm C})]/[n({\rm O})(1 - f_{\rm dpl}({\rm O})]=0.6$, described in text.
\end{tablenotes}
\end{table}

For the \cloudy models with $Z/$Z$_{\odot}=1$, we adopt the ISM abundances derived by \citet{Savage1977} and the ISM oxygen abundance by \citet{Meyer1998}, together with their metal depletion onto dust grains (see Table~\ref{tab:abund:deplet}), the typical ISM grains, and the C/O ratio parameterized by the metallicity.
Several observations implied that the C/O ratio depends on the metallicity \citep[e.g.,][]{Garnett1995,Garnett1999}. Considering the C/O measurements by \citet{Garnett1999}, 
the C/O ratio correlates with the O/H abundance as $\log {\rm C}/{\rm O} = 3.84+ 1.16 \log{\rm O}/{\rm H}$, so
we get ${\rm C}/{\rm O}=0.16$ for the Small Magellanic Cloud (SMC; $\log {\rm O}/{\rm H}=-4$, i.e., $Z/$Z$_{\odot} \sim 1/5$), ${\rm C}/{\rm O}=0.46$ for the Large Magellanic Cloud (LMC; $\log {\rm O}/{\rm H}=-3.6$, i.e., $Z/$Z$_{\odot} \sim 1/2$), 
and ${\rm C}/{\rm O}=0.6$ for the local Galactic ISM ($\log {\rm O}/{\rm H}=-3.5$, i.e., $Z/$Z$_{\odot} \sim 1$).
The carbon abundance is configured to yield the depleted C/O ratio associated with the relevant metallicities (see Table~\ref{tab:abund:deplet}). 
We apply the \cloudy ``metals'' command to the models with the metallicities $Z/$Z$_{\odot}=0.5$, $0.25$, and $0.125$ to scale down the baseline solar abundances of elements heavier than helium.
The same metal depletion and adjusted C/O ratios are also incorporated into \cloudy models used by \maihem\
for initializing CIE states of the ejected gas, as well as for metal
species used in the NEI calculations by our \maihem\ simulations. 

For our \cloudy models, we also include the typical ISM grains with a dust-to-metal mass ratio of $M_{\rm d}/M_{\rm Z}=0.2$, 
which is typically associated with evolved galaxies \citep{DeVis2019}.
Previously, the dust-to-gas versus metallicity correlation \citep[see e.g.][]{Hirashita1999,Edmunds2001} corresponds to $M_{\rm d}/M_{\rm Z} \sim 0.5$. However, the recent study by \citet{DeVis2019} suggested that evolved galaxies with metallicities above $\log {\rm O}/{\rm H}=-3.8$ have a more or less constant dust-to-metal ratio of $M_{\rm d}/M_{\rm Z} \approx 0.214$, while significantly lower values are expected for unevolved galaxies ($\log {\rm O}/{\rm H} < -3.8$). As we do not consider implications of dust grains in this paper, we assume the standard ISM grains having a constant ratio of $M_{\rm d}/M_{\rm Z}=0.2$. The dust grains along with depleted metallic species are included in \cloudy models for initializing CIE states of the outflow in \maihem, but not in the NEI module. 

\subsection{Population Synthesis Models}
\label{cooling:starburst99}

The spectral energy distribution (SED) of the radiation emitted from stars in the SSC plays an important role in producing the ionization states and emission lines. For the ionizing input of our photoionization models, we employ the latest version of the evolutionary synthesis code Starburst99 \citep[][]{Leitherer1999,Leitherer2014} that was optimized for stellar population with various ages \citep{Vazquez2005}, as well as extended to include rotational mixing effects \citep{Levesque2012,Leitherer2014}. 

We assume the same fixed SSC mass of $2.05\times10^6$\,M$_{\odot}$ at all 
metallicities.  This $M_\star$ corresponds to $\dot{M}_{\rm sc} = 10^{-2}$\,M$_{\odot}$\,yr$^{-1}$
for the fiducial Starburst99 model with $Z/$Z$_{\odot}=1$.
We use an initial mass function (IMF) 
having a power-law $dN/dm \varpropto m^{-\alpha}$ with the Salpeter value $\alpha= 2.35$ \citep{Salpeter1955}, 
over a mass range of 0.5 to 150 M$_{\odot}$. We consider the 
Geneva stellar evolution with stellar rotation (Geneva-Rot) implemented 
by \citet{Levesque2012} and \citet{Leitherer2014},
which incorporates stellar population grids of 
\citet{Ekstroem2012,Georgy2012}. 
and \citet{Georgy2013}. 
We also employ  the stellar wind models of
\citet{Maeder1990} and the extended Pauldrach/Hillier atmosphere models \citep{Hillier1998,Pauldrach2001}, which are appropriate for the O-type stellar atmospheres. The Starburst99 high-resolution spectra are built using 
the evolutionary population synthesis by \citet{Martins2005}. 
The different evolution and atmosphere models have varying
metallicity, and we use the closest match to the metallicities in our
simulations, as summarized in Table~\ref{tab:sim:param}.

For our stellar population models, we use an age of 1 Myr,
corresponding to the fiducial age of our hydrodynamic simulations. 
Table~\ref{tab:sim:param} also lists the input parameters used in our Starburst99 population synthesis calculations that correspond to
the fiducial $M_* = 2.05\times 10^6$\,M$_{\odot}$ at $Z/$Z$_{\odot} = 1$. 
The stellar population models yield the ionizing luminosities and SED
continua that are used as inputs by our \cloudy photoionization
models for the adopted fixed $M_\star$.
Table~\ref{tab:sim:param} presents the results from the stellar population models generated by Starburst99:
the total luminosity ($L_{\rm tot}$), 
the fraction of ionizing photons in the H$^{+}$ continuum relative to the total luminosity ($f($H$^{+})$),
and the ionizing luminosity ($L_{\rm ion}$). 
The ionizing luminosity, which is calculated according to the H$^{+}$ fraction and the total luminosity, together with 
the synthetic stellar spectrum (SED) are used as inputs in our \cloudy photoionization modeling. 
Decreasing the metallicity $Z/$Z$_{\odot}$ with the fixed total stellar mass slightly increases the total luminosity,
since metal-poor stars are somewhat hotter and more luminous.

\begin{figure}
\begin{interactive}{js}{figure2.tar.gz}
\includegraphics[width=0.42\textwidth, trim = 0 0 0 0, clip, angle=0]{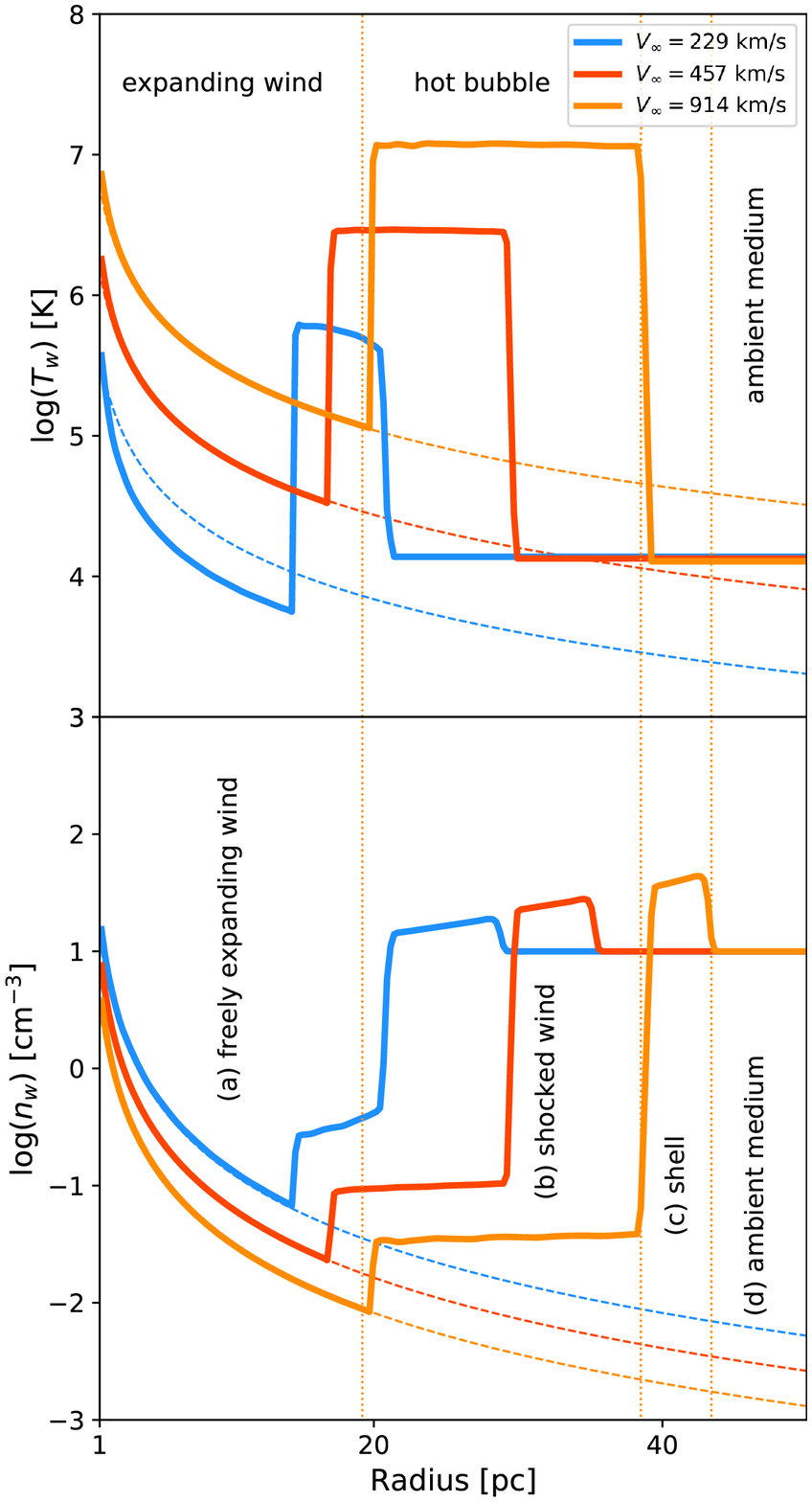}
\end{interactive}
\caption{From top to bottom: temperature and density profiles of
  representative \maihem\ models on a logarithmic scale with 
  ambient density of $n_{\rm amb}=10$ cm$^{-3}$, and 
wind terminal speeds of $V_{\infty}=229$ (blue), $457$ (red), and
  $914$\,km\,s$^{-1}$ (orange solid line). The metallicity is $Z/$Z$_{\odot}=0.5$ and 
mass-loss rate $\dot{M}_{\rm sc}=6.07 \times 10^{-4}$ M$_{\odot}$\,yr$^{-1}$.
The SSC has radius $R_{\rm sc}=1$ pc and age $t=1$ Myr. 
The different regions separated by dotted lines are labeled for $V_{\infty}=914$\,km\,s$^{-1}$:  (a) freely expanding wind, 
(b) shocked-wind, (c) shell, and (d) ambient medium.
The adiabatic temperature and density profile for each model are shown by dashed lines.
The winds are in the CB (229\,km\,s$^{-1}$) and AB ($457$, $914$\,km\,s$^{-1}$) modes, described in \S\,\ref{cooling:hydrodynamics:results}. 
The plots for the entire model grid (64 images) are available in the interactive figure 
in the online journal.
\label{fig:temp:dens:profiles} }
\end{figure}

\begin{figure}
\begin{center}
\begin{interactive}{js}{figure3.tar.gz}
\includegraphics[width=0.42\textwidth, trim = 0 0 0 0, clip, angle=0]{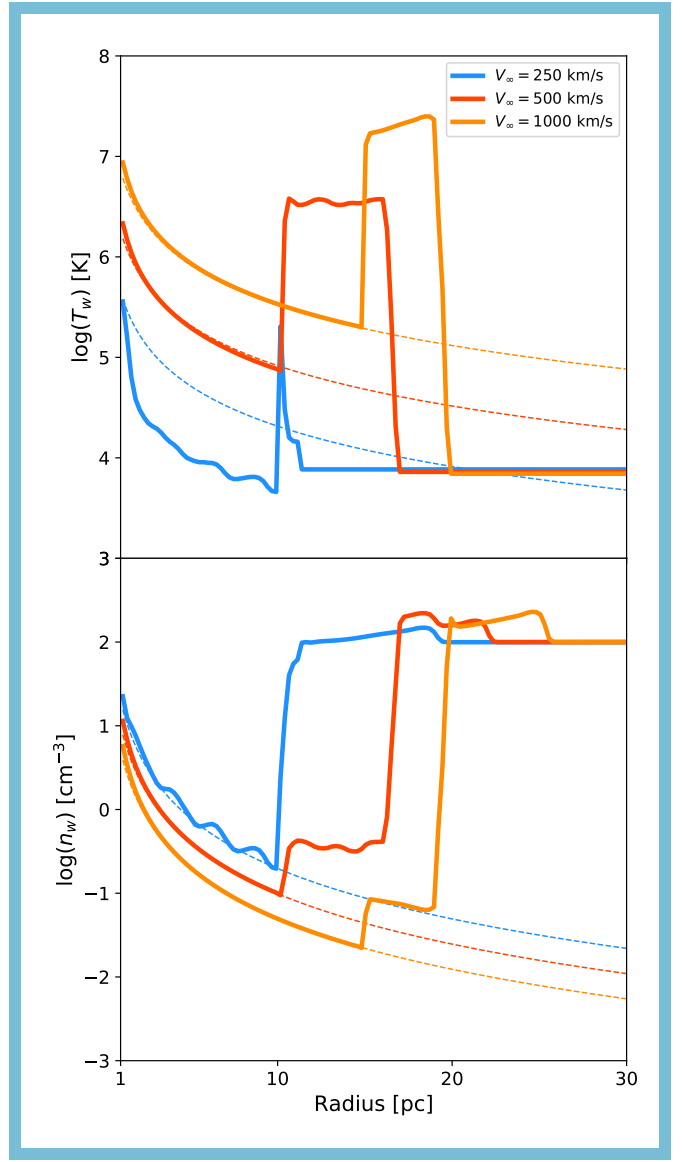}
\end{interactive}
\end{center}
\caption{Similar to Figure~\ref{fig:temp:dens:profiles}, but showing a
  catastrophic cooling model 
with ambient density of $n_{\rm amb}=100$ cm$^{-3}$, 
and wind terminal speed of $V_{\infty}=250\ \rm km\ s^{-1}$ (blue); together with
models for $V_\infty=500$ (red), and $1000$\,km\,s$^{-1}$ (orange solid line).
The metallicity is $Z/$Z$_{\odot}=1$ and 
mass-loss rate $\dot{M}_{\rm sc}=10^{-3}$\,M$_{\odot}$\,yr$^{-1}$. 
The adiabatic temperature and density profile for each model are shown by dashed lines. 
The winds are in the CC (250\,km\,s$^{-1}$) and AB ($500$, $1000$\,km\,s$^{-1}$) modes, described in \S\,\ref{cooling:hydrodynamics:results}. 
The plots for the entire model grid (64 videos) are available in the interactive animation in the online journal. 
\label{fig:temp:dens:profiles2} 
}
\end{figure}

\begin{figure*}
\begin{center}
\includegraphics[width=0.8\textwidth, trim = 0 0 0 0, clip, angle=0]{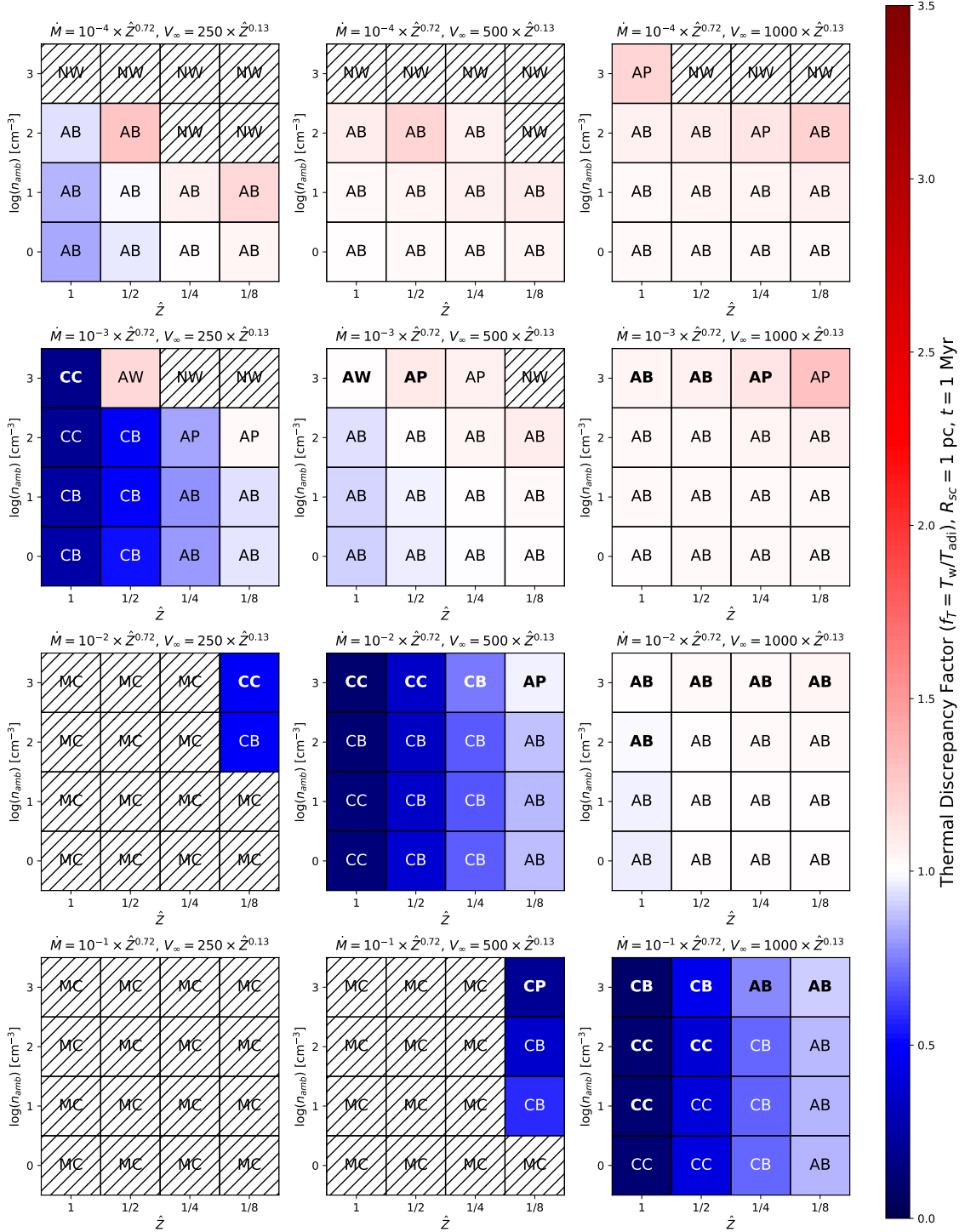}
\end{center}
\caption{The temperature discrepancy factor ($f_T \equiv T_{\rm w} / T_{\rm adi}$) as a function of
the metallicity ($\hat{Z} \equiv Z/$Z$_{\odot}=1$, 0.5, 0.25, and 0.125), 
the mass-loss rate ($\log \dot{M}_{\rm sc}= -1$, $-2$, $-3$, and $-4$\,M$_{\odot}$\,yr$^{-1}$ at $Z=$\,Z$_{\odot}$) coupled to the metallicity ($\dot{M}_{\rm sc}\varpropto Z^{0.72}$; see Table~\ref{tab:sim:param}),
and the wind terminal velocity ($V_{\infty}=250$, 500, and 1000 km\,s$^{-1}$ at $Z=$\,Z$_{\odot}$)
coupled to the metallicity ($V_{\infty}\varpropto Z^{0.13}$; see Table~\ref{tab:sim:param})
, and the ambient density ($\log n_{\rm amb}=0$, 1, 2, and 3 \,cm$^{-3}$) computed by the \maihem 
hydrodynamic simulations
with the cluster radius of $R_{\rm sc} =1$\,pc and the fixed cluster mass of $2.05 \times 10^6$\,M$_{\odot}$. 
The current age of the hydrodynamic iteration is 1 Myr.
The wind classification modes based on the temperature profile are presented, namely 
the 
adiabatic wind (AW),  
adiabatic bubble (AB), adiabatic, pressure-confined (AP), 
catastrophic cooling (CC),  catastrophic cooling bubble (CB), 
catastrophic cooling, pressure-confined (CP), 
no expanding wind (NW), and momentum-conserving (MC), 
while the wind modes of the optically thick models are denoted by the bold font. 
\label{fig:cooling:temp:fact} 
}
\end{figure*}

\begin{figure*}
\begin{center}
\includegraphics[width=0.8\textwidth, trim = 0 0 0 0, clip, angle=0]{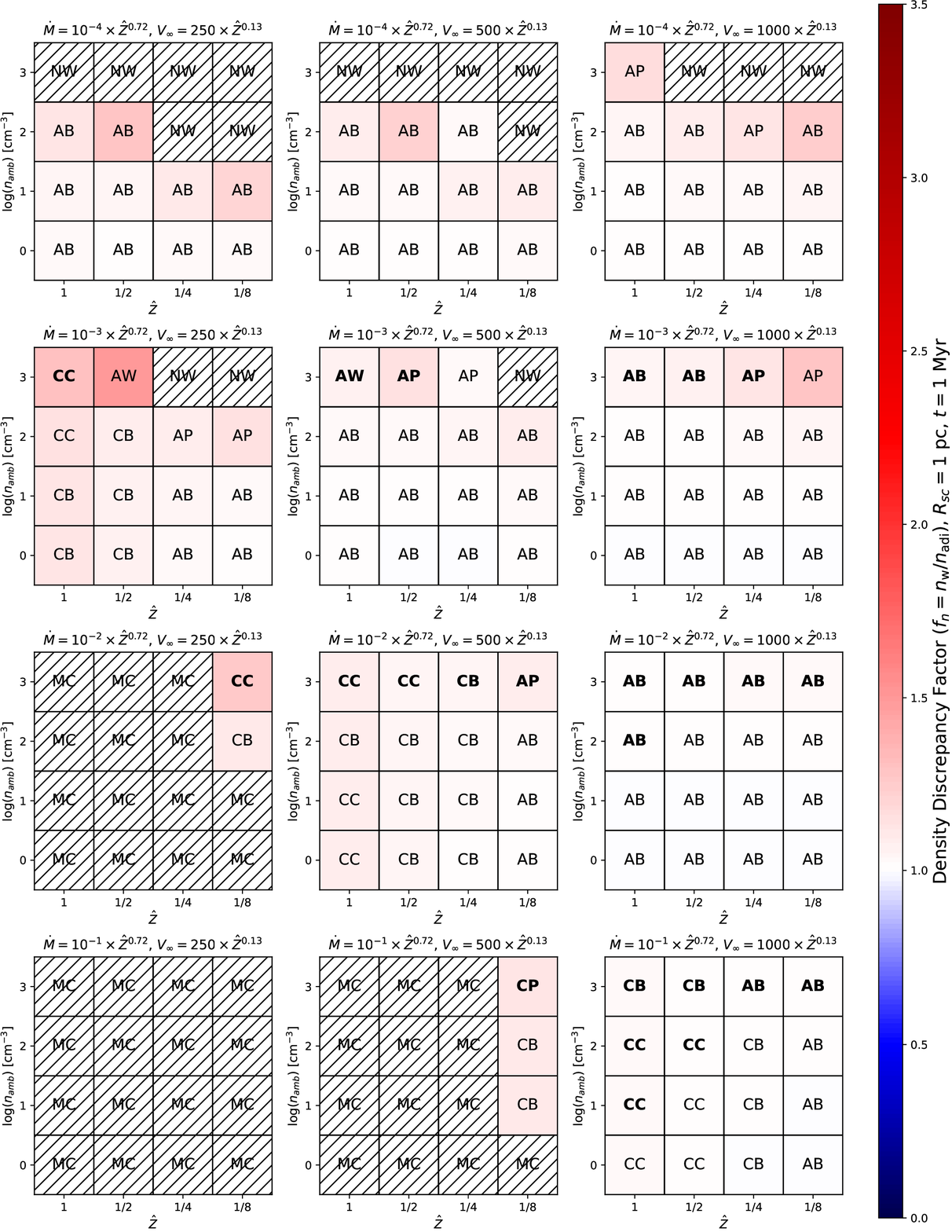}
\end{center}
\caption{The same as Figure~\ref{fig:cooling:temp:fact}, but for the density discrepancy factor ($f_n \equiv n_{\rm w} / n_{\rm adi}$)
\label{fig:cooling:dens:fact}
}
\end{figure*}


\section{Hydrodynamic Results}
\label{cooling:hydrodynamics:results}

In this section, we present the results of our hydrodynamic simulations performed by \maihem using the parameters listed in Table~\ref{tab:sim:param}. 

To illustrate different features of the temperature and density
profiles in the adiabatic and radiative cooling modes, in
Figure~\ref{fig:temp:dens:profiles} we plot the temperature $T_w$ and density $n_w$ as a
function of radius $r$.  
Considering the hydrodynamic simulations displayed as an example figure at $t=1$ Myr with the metallicity $Z/$Z$_{\odot}=0.5$, mass-loss rate $\dot{M}_{\rm sc}=6.07 \times 10^{-4}$ M$_{\odot}$\,yr$^{-1}$, cluster radius $R_{\rm sc}=1$ pc, ambient density $n_{\rm amb}=10$ cm$^{-3}$, and ambient temperature $T_{\rm amb} \sim 1.4 \times 10^4$\,K determined by \cloudy: there are two with negligible radiative cooling ($V_{\infty}=457$ and $914$\,km\,s$^{-1}$), and one with considerable radiative cooling ($V_{\infty}=229$\,km\,s$^{-1}$).  
The hydrodynamic models ($V_{\infty}=457$ and $914$\,km\,s$^{-1}$) 
without radiative cooling are not continuous freely expanding CC85 winds due to the presence of a shocked-wind region described by \citet{Weaver1977}, but possess the CC85 adiabatic solution before the bubble in their temperature
and density
profiles. The hydrodynamic model ($V_{\infty}=229$\,km\,s$^{-1}$) with
strongly radiative cooling does not follow the CC85 adiabatic cooling
flow before the shell, and it has a radiative solution more similar to
that described by \citet{Silich2004}. 

The adiabatic cooling mode in $V_{\infty}=914$ km\,s$^{-1}$ shown in the example image of Figure~\ref{fig:temp:dens:profiles} has four different regions described in \citet{Weaver1977}: (a) freely expanding supersonic wind, (b) shocked wind region, (c) a narrow shell of shocked ambient medium, and
(d) the ambient medium. We see that the temperature and density profiles in the region (a) follows the prediction by the CC85 analytic adiabatic solution (dashed line). As the wind propagates into the ambient medium,
a hot bubble of shocked wind results, with
sharply increased temperature. This leads to the formation of a dense shell due to the confrontation with the ambient medium. The shell has a density higher than the ambient density.

The non-adiabatic, radiative cooling mode is shown in an example figure of Figure~\ref{fig:temp:dens:profiles2}:  this model  ($V_{\infty}=250$ km\,s$^{-1}$) 
has all the regions described in the adiabatic cooling mode, 
except for the shocked wind region (b) corresponding to the hot bubble. 
The animated version of this figure shows the evolution for $t=0.1,0.2,\ldots,1$ Myr (available in the online
journal).\footnote{The interactive animation for all the models (64
  videos) is archived on  
Zenodo (doi:\href{https://doi.org/10.5281/zenodo.4989577}{10.5281/zenodo.4989577}).} The strong cooling effects suppress the formation of a powerful shocked wind region and an intense temperature bubble.
Moreover, the temperature of the freely expanding supersonic wind does not follow the analytic adiabatic solution (dashed line). 
It can be seen that
 radiative cooling strongly reduces
the temperature of the freely expanding supersonic wind that leads to the deposition of much lower kinetic and thermal energy in 
the surrounding ambient medium.
Thus, we identify three regions in the strongly radiative cooling (catastrophic cooling) mode: freely expanding supersonic wind, a broad shell with density slightly higher than the ambient density, and the surrounding medium.  This model corresponds to a momentum-driven shell.

Figure~\ref{fig:temp:dens:profiles2} depicts adiabatic cooling with $V_{\infty}=1000$ km\,s$^{-1}$ that does not develop any bubble at ages $t\leq 0.9$ Myr (see its animation in Supplementary Material). 
However, this model generates a bubble at $t=1$ Myr, 
so the hot bubble characteristic of the adiabatic solution
can take time to develop. 
In our grid, the model for
$V_{\infty}=500$\,km\,s$^{-1}$, $n_{\rm amb}=1000$\,cm$^{-3}$, $Z/$Z$_{\odot}=1$ at
$\dot{M}_{\rm sc}=10^{-3}$ M$_{\odot}$\,yr$^{-1}$ generates bubbles at 1.1\,Myr that is greater than the 1 Myr age adopted for all the models. 
It is important to note that {\it lack of a hot bubble
in young objects does not necessarily indicate that the system is strongly cooling} or otherwise not adiabatic. 
Conversely, the hot, shocked wind region can also be present in our simulations at $t=1$ Myr with radiative cooling (e.g. $V_{\infty}=229$ km\,s$^{-1}$ in Figure~\ref{fig:temp:dens:profiles}), whose freely expanding wind experiences mildly cooling, but not strong enough to suppress the shocked-wind and hot bubble.

The temperature and density profiles of the superwinds simulated by our hydrodynamic models are therefore classified according to
the presence of adiabatic/radiative cooling 
and the bubble as follows:
\begin{enumerate}
\item adiabatic wind (AW) mode: The temperature profile is roughly of freely expanding supersonic winds (CC85) that
closely follows the adiabatic solution, and it has not yet produced any bubble, which may develop at a later time. 
The temperature $T_{\rm w}$ of the expanding wind (region a in Figure~\ref{fig:temp:dens:profiles}) produced by \maihem is within 0.75 and 1.25 times the temperature $T_{\rm adi}$ analytically derived from the adiabatic solution.

\item adiabatic bubble (AB) mode: This is the classic bubble model \citep{Weaver1977} where the temperature profile is roughly of adiabatic winds (slightly or without radiative cooling) surrounded by a shocked-wind region (hot bubble; with a thickness $> 0.73$\,pc and a mean temperature $> 10^5$\,K).
The temperature profile closely follows the adiabatic solution of freely expanding supersonic winds (CC85). 
The temperature $T_{\rm w}$ of the expanding wind computed by \maihem\  is within 0.75 and 1.25 of 
the analytic adiabatic temperature $T_{\rm adi}$ over the expanding wind region.
The hot bubble (region b in Figure~\ref{fig:temp:dens:profiles}) drives the shell expansion.

\item  adiabatic, pressure-confined (AP) mode: 
The expanding wind region is adiabatic, but the bubble expansion is stalled by high ambient pressure from the dense environment.  This corresponds to the pressure-confined model predicted by \citet{Silich2007} \citep[also see][]{Koo1992}. 
The temperature profile is roughly of freely expanding supersonic winds (CC85).
The expanding wind temperature $T_{\rm w}$ computed by \maihem\ is between 0.75 and 1.25 times the temperature $T_{\rm adi}$
analytically derived from the adiabatic solution. 

\item catastrophic cooling (CC) mode: The temperature profile is of freely expanding winds with strongly radiative cooling \citep{Silich2003,Silich2004} without any noticeable shock regions and weak/no bubble. The expanding wind temperature $T_{\rm w}$ of the freely expanding wind (region a) simulated by \maihem is at least 25\% lower than the temperature $T_{\rm adi}$ analytically derived from the adiabatic solution.

\item catastrophic cooling bubble (CB) mode: The temperature profile has strongly radiative cooling, but still has the bubble and shocked-wind regions. The expanding wind temperature $T_{\rm w}$ of the region (a) is at least 25\% below the adiabatic temperature $T_{\rm adi}$ from the analytic solution. Additionally, when the bubble is confined by the high pressure from the ambient medium, we classify it as the cooling, pressure-confined (CP) mode.

\item no wind (NW) mode:  The mass-loss rate is extremely low
(e.g. $10^{-4}$ M$_{\odot}$\,yr$^{-1}$) and the ambient medium is very
dense (e.g. $n=10^3$ cm$^{-3}$ here), so freely expanding supersonic
wind is completely inhibited.  The wind solution does not meet the
outflow pressure criterion of \citet{Canto2000} (see below).

\item momentum-conserving (MC) mode:
The mass-loss rate is high and the velocity is low, causing
catastrophic cooling so that the wind is suppressed.
We classify as MC mode those models having
a mean initial temperature in region (a) that is a factor of
3 below the expected mean adiabatic temperature
within a distance  of 1\,pc from the cluster boundary.
\end{enumerate}

\begin{figure*}
\begin{interactive}{js}{figure6.tar.gz}
\includegraphics[width=0.8\textwidth, trim = 0 0 0 0, clip, angle=270,origin=rb]{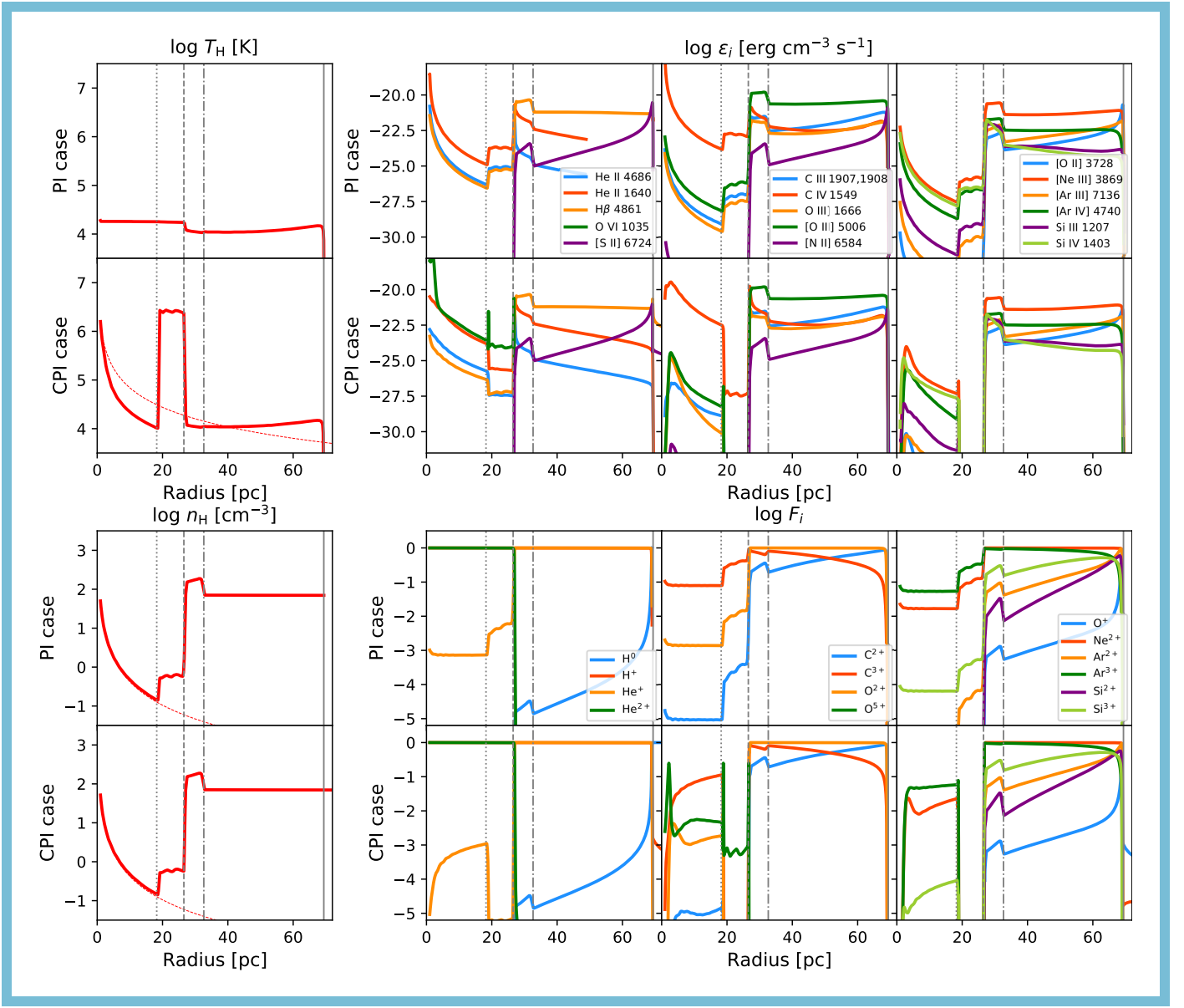}
\end{interactive}
\caption{\textit{Top Panels}: The hydrogen temperature profiles ($T_{\rm H}$\,[K]; left panels) along with the adiabatic prediction (red dashed line) in the CPI case, and
  the line emissivities ($\epsilon_{i}$\,[erg\,cm$^{-3}$\,s$^{-1}$]; right
  panels) of the superwind models on a logarithmic scale, from top to
  bottom, in the 
  PI (pure photoionization) and CPI (photoionization $+$ hydrodynamic collisional ionization) cases
  (\S\,\ref{cooling:photoionization}), for wind terminal
  speed $V_{\infty}=457$\,km\,s$^{-1}$, metallicity
  $\hat{Z}\equiv Z/$Z$_{\odot}=0.5$, mass-loss rate $\dot{M}_{\rm sc}= 0.607 \times 10^{-2}$
  M$_{\odot}$\,yr$^{-1}$, cluster radius $R_{\rm sc}=1$ pc, and age
  $t=1$ Myr, surrounded by the ambient medium with density $n_{\rm
    amb}=100$ cm$^{-3}$ and temperature $T_{\rm amb}$ determined by
  \cloudy, producing a wind model in the CB mode (described in \S\,\ref{cooling:hydrodynamics:results}).
   \textit{Bottom Panels}: The hydrogen density profiles ($n_{\rm H}$\,[cm$^{-3}$]; left panels) along with the adiabatic prediction (red dashed lines), and the ionic fractions ($F_{i}$; right panels) on a logarithmic scale for the 
PI and CPI cases.
  The start and end of the hot bubble (region b), the end of
  the shell (region c), and the Str\"{o}mgren radius are shown by
  dotted, dashed, dash-dotted, and solid lines (gray color),
  respectively.  
  The plots for all the models (192 images) are available in the interactive figure in the online journal. 
\label{fig:CIE:emissivity}}
\end{figure*}

The classifications fall into three categories:  adiabatic and
quasi-adiabatic (AW, AB, AP), radiatively cooling (CC, CB, CP), and
suppressed superwinds (NW, MC).  The suppressed superwinds are the
extreme cases.  The NW mode represents the case where the 
mass-loss rate is too low relative to the ambient density, and a wind
cannot launch.
As mentioned by \citet{Canto2000}, the existence of the supersonic
wind depends on the following outflow pressure: 
\begin{equation}
P_{\rm out} \equiv \frac{1}{\gamma} \left( \frac{\gamma-1}{\gamma+1} \right)^{\frac{1}{2}} 
\left( \frac{\gamma+1}{2} \right)^{\frac{\gamma}{\gamma-1}}
 \frac{ \dot{M}_{\rm sc} V_{\infty}}{ 4 \pi  R_{\rm sc}^2 },
\label{eq:13}%
\end{equation}
To have the supersonic solution, it is necessary to have the outflow
pressure $P_{\rm out}$ higher than the ambient thermal pressure
$P_{\rm amb}$ in the surrounding ambient medium. In the ideal gas
condition, the ambient thermal pressure $P_{\rm amb}$ is a function of
the ambient temperature and density as $P_{\rm amb} =  k_{\rm B}
T_{\rm amb} \rho_{\rm amb}/ \mu$, where $k_{\rm B}$ is Boltzmann's
constant and $\mu$ is the mean mass per particle of the gas. This
provides a physical explanation for the dependence of the wind
existence upon the mass-loss rate, the wind terminal velocity, and the
ambient density, for a given cluster radius ($R_{\rm sc}=1$\,pc) and
ambient temperature.
Outflow with very low mass-loss rate and/or terminal velocity does not
support the supersonic solution, so there is no freely expanding
supersonic wind. From Eq. (\ref{eq:13}), it can be seen that
supersonic conditions also depend on the cluster radius ($R_{\rm
  sc}$), so at a very large value of $R_{\rm sc}$, depending on
$\dot{M}_{\rm sc}$  and $ V_{\infty}$, the existence of freely
expanding supersonic winds is also impossible.

Superwinds are also suppressed by the inverse situation where the
mass-loss rate is large and the velocity low, causing strong
catastrophic cooling.
The MC mode corresponds to conditions like these, where
there is no significant thermal energy
contributing to the outflow, and it roughly proceeds in a momentum-conserving
mode only \citep[see e.g.][]{Ostriker1988,Koo1992,Koo1992a}.
For such models with high mass-loss rates and low wind speeds
(e.g. $V_{\infty}=250$ km\,s$^{-1}$ for $10^{-2}$
M$_{\odot}$\,yr$^{-1}$), it might often be the case that a supersonic
wind cannot be launched at the injection radius $R_{\rm,sc}$ \citep[e.g.,][]{Silich2018}. 
This could arise in situations such as those studied by
\citet{Silich2018} where hot shocked winds from individual stars
strongly cool down inside the cluster before a superwind can be launched
into the ambient medium. Previously, it was shown that high
mass-loading can contribute to strong radiative cooling in the hot
shocked wind \citep{Lochhaas2018,Silich2020} that may also suppress or
delay the development of a star cluster superwind \citep{Silich2020}.

The parameter space for the different regimes can be seen in 
Figure~\ref{fig:cooling:temp:fact}.  This shows the temperature discrepancy factor ($f_T \equiv T_{\rm w} / T_{\rm adi}$) defined as the mean value of 
the expanding wind temperature $T_{\rm w}$ over the predicted analytic
adiabatic temperature $T_{\rm adi}\varpropto r^{-4/3}$ for the freely
expanding supersonic wind, region (a). A value of $f_T$ lower than 0.75 is associated with strongly radiative cooling (CC and CB modes). 
A value for the temperature discrepancy factor of $\sim 1$ generally corresponds
to the adiabatic cooling mode with $0.75<f_T <1.25$ (Model AB). It can be seen
that, as expected, an increase in the mass-loss rate decreases the
temperature discrepancy factor, because the higher density
increases cooling. Increasing the metallicity also reduces $f_T$, thus
also increasing radiative cooling \citep[e.g.,][]{Silich2004}.  The MC models represent the
extreme case, where there is no thermal energy contributing to the outflow.
We also see that for a given $\dot{M}_{\rm sc}$, catastrophic cooling
occurs for lower outflow velocities.
While strong cooling is also promoted by higher metallicity,
Figure~\ref{fig:cooling:temp:fact} shows that $\dot{M}_{\rm sc}$
and  $V_{\infty}$ are more important over the modeled parameter
space.

The pressure-confined models (AP) tend to have higher temperatures, as
expected since the adiabatic expansion is reduced, therefore elevating
the temperatures in the interior regions according to the ideal gas law.
Two models have $f_T > 1.25$.  They are identified
in Figure~\ref{fig:cooling:temp:fact} as 
$\dot{M}_{\rm sc}=10^{-4}$ M$_{\odot}$\,yr$^{-1}$,  $V_{\infty}=1000$\,km\,s$^{-1}$, $n_{\rm amb}=1000$ cm$^{-3}$, $Z/$Z$_{\odot}=1$; and
$\dot{M}_{\rm sc}=0.224 \times 10^{-3}$ M$_{\odot}$\,yr$^{-1}$,  $V_{\infty}=736$\,km\,s$^{-1}$,\ $n_{\rm amb}=1000$ cm$^{-3}$, $Z/$Z$_{\odot}=0.125$. 
These are models at the high-density limit, where bubble growth is
inhibited the most.

As shown in Figure~\ref{fig:cooling:dens:fact}, the density
discrepancy factor ($f_n \equiv n_{\rm w} / n_{\rm adi}$) defined as
the mean value of the expanding wind density $n_{\rm w}$ over the
adiabatic density $n_{\rm adi}$ is computed for the freely expanding
supersonic wind (region a).
It can be seen that the density profiles mostly follow the adiabatic solution.
However, the regions with strong cooling (CC, CB)
have slightly enhanced densities, as expected.  
High densities could also lead to high ambient pressures, 
which also lead to slower bubble growth, sometimes resulting in the adiabatic,
pressure-confined (AP) models.  

We also see that the wind mechanical power $\frac{1}{2} \dot{M}_{\rm sc} V^2_{\infty}$
does not itself determine the outflow's status with respect to
catastrophic cooling; some models with the weakest wind power (upper
left group in Figure~\ref{fig:cooling:temp:fact}) are still
adiabatic.  Rather, $\dot{M}_{\rm sc}$ and $V_{\infty}$
are independently the critical parameters in establishing strong cooling.
Moreover, we see that 
there are more catastrophic cooling models that are
still capable of supporting hot bubbles (CB) than those without (CC).
Thus, {\it the presence of a hot, X-ray superbubble does not necessarily indicate
adiabatic feedback.}

Figure~\ref{fig:cooling:temp:fact} shows that for the models
with fiducial metallicity scalings for the effective mass-loss rate and
wind velocity ($\dot{M}_{\rm sc}=10^{-2}$ M$_{\odot}$\,yr$^{-1}$ and
$V_{\infty}=1000$\,\ km\,s$^{-1}$), essentially all the models show
conventional adiabatic feedback.  Reducing  $\dot{M}_{\rm sc}$ does
not change this scenario until the outflow simply cannot penetrate
the ambient density.  However, reducing the wind velocity, or otherwise
reducing the kinetic heating efficiency, destabilizes
these models to catastrophic cooling.  Reducing the wind density can
mitigate this effect, but only to a point, as seen in these model
grids.  In practice, regions that tend to be radiation dominated are
probably more likely to be mass-loaded due to 
photoevaporation.  A smaller outflow region, which keeps the gas
close to the SSC, also promotes these effects.  
On the other hand, if catastrophic cooling
takes place inside the launch radius \citep[e.g.,][]{Silich2018,
Silich2020}, it could result in a very small effective
$\dot{M}_{\rm sc}\sim 10^{-4}\ \rm M_\odot\ yr^{-1}$.
This could potentially generate the parameter space of models in that
regime, which are otherwise unrealistic for stellar wind outflows.


\section{Predicted Emission-Line Ratios}
\label{cooling:cloudy:results}

In this section, we present emission line emissivities and luminosities determined
using \cloudy, for the two different models in
\S\,\ref{cooling:hydrodynamics:results}, namely, PI (pure
photoionization) and CPI (photoionization $+$ hydrodynamic collisional ionization). 

Figure~\ref{fig:CIE:emissivity} (top panels) shows the emissivities calculated for various
emission lines as a function of radius for 
our photoionization models with the two different ionization schemes (PI and CPI; right panels). 
The PI emissivity profiles, which were made using the
Starburst99 SED and luminosity applied to the neutral density profile
from the hydrodynamic simulation, are shown in the top panel. The
bottom panel presents the emissivities from the CPI model
built using both the Starburst99 SED ionizing source and hydrodynamic ionization. 
Similarly, the ionic fractions for our photoionization models with these two different (PI and CPI) models are presented in Figure~\ref{fig:CIE:emissivity} (bottom panels).

The CPI models are used to calculate line emissivities in what follows.
We classify our photoionization models according to the nebular (H$^+$) optical depth 
of the swept-up shell. Optically thick models are shown with bold-face 
in Figure~\ref{fig:cooling:temp:fact}.
Models are optically thin if the ambient medium
beyond the shell (region c in Figure~\ref{fig:temp:dens:profiles}) is photoionized to produce H$^+$ by the SSC. 
We see that the optically thick models are those in the densest
($n_{\rm amb} \gtrsim 10^2$) ambient media, for which dense, optically
thick swept-up shells develop more quickly.
However, we caution that our 1-D models do not account for shell
clumping which can greatly alter the escape fraction of ionizing radiation.
We also caution that \maihem\ does not implement radiative transfer, 
which may be a significant effect 
in high-density models. 
Thus, the exact boundary of the ionized edge is approximate,
and the ionization structure of the interface region between the
optically thin and thick conditions are also approximated. 
Thus for the optically thick \cloudy\ 
models, we adopt the radial temperature profile from the PI model 
from the radius at
which the shell becomes isothermal according to the \maihem simulation,
roughly 1--2 pc outward from the inner boundary of the dense shell (see
Section \ref{cooling:photoionization}). 

Observations of distant, luminous complexes may be biased  
toward the central, high surface-brightness regions, thereby
approximating density bounded observations transverse to the line of sight.
To allow for the calculation of partially and fully radiation-bounded models, the total luminosity $L_{\lambda}$ 
of each emission line at wavelength $\lambda$ is calculated from its volume emissivity $\epsilon_{\lambda} (r)$ as follows (see Appendix~\ref{appendix:a}):
\begin{equation}
L_{\lambda} = 4 \pi \int^{R_{\rm aper}}_{R=0}  \left[ \int^{R_{\rm max}}_{r=R}  \frac{\epsilon_{\lambda} (r)}{ \sqrt{r^2 - R^2} }  r dr \right] R d R,
\label{eq:20}%
\end{equation}
where $r$ is the radial distance from the center, $R$ the projected distance from the center, $R_{\rm aper}$ the radius of the circular projected aperture within the object observed, and $R_{\rm max}$ the maximum radius of the object in the line of sight. Thus, for a fully ``density-bounded'' model, we set $R_{\rm max}=R_{\rm shell}$, i.e., the maximum radius of the dense shell (region c in Figure~\ref{fig:temp:dens:profiles}), and  $R_{\rm aper}=R_{\rm shell}$.  For a fully ``radiation-bounded'' \ionic{H}{ii} region, $R_{\rm max}$ and $R_{\rm aper}$ are set to the Str\"{o}mgren radius $R_{\rm Str}$ predicted by \cloudy. Setting $R_{\rm aper}=R_{\rm shell}$ and $R_{\rm max}=R_{\rm Str}$ is radiation bounded in the line of sight, but otherwise density bounded.  This geometry is described as a ``partially density-bounded'' model in what follows.
Appendix~\ref{appendix:a} fully describes the calculations based on this geometry.

Table~\ref{tab:cloudy:output} partially lists the total luminosities derived from emissivities calculated by \cloudy for radiation-bounded, partially density-bounded, and density-bounded models with the two different cases (PI and CPI). The tables for all the photoionization calculations are presented in the machine-readable format in Appendix~\ref{appendix:b}. 

\begin{figure*}
\begin{center}
\includegraphics[width=0.75\textwidth, trim = 0 0 0 0, clip, angle=0,origin=rb]{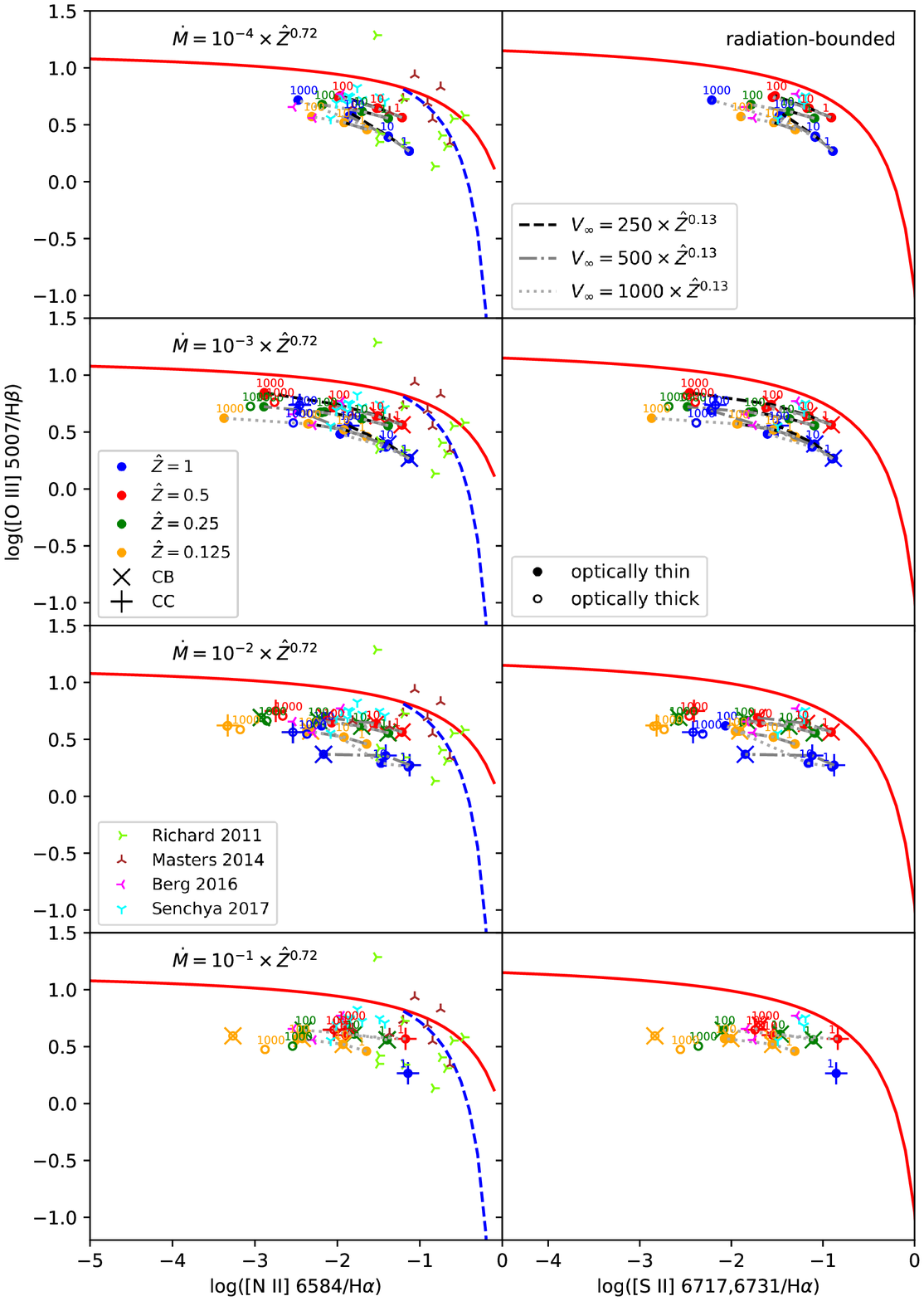}
\end{center}
\caption{The optical diagnostic BPT diagrams plotted \foiii $\lambda$5007/\hb versus \fnii $\lambda$6584/\ha (left) and  \foiii $\lambda$5007/\hb versus \fsii $\lambda\lambda$6717,6731/\ha (right panels) 
for the fully radiation-bounded models
with mass-loss rates $\log \dot{M}_{\rm sc} =-4$, $-3$, $-2$, and $-1$ M$_{\odot}/$yr (from top to bottom), 
ambient densities $n_{\rm amb}= 1$, 10, $10^2$, and $10^3$\,cm$^{-3}$ (labeled on plots), 
metallicities $\hat{Z} \equiv Z/$Z$_{\odot}=1$ (blue), $0.5$ (red), $0.25$ (green), and $0.125$ (yellow color), and wind terminal velocities $V_{\infty}=250$ (dashed), $500$ (dash-dotted), and $1000$ km\,s$^{-1}$ (dotted lines).
For the sub-solar models, we use the solar model parameters scaled as $\dot{M}_{\rm sc} \varpropto Z^{0.72}$ and $V_{\infty} \varpropto Z^{0.13}$.
The optically thin and thick models are plotted by filled and empty
circles, respectively.
Line emissivities for optically thick models are somewhat uncertain.
The wind catastrophic cooling (CC) and catastrophic cooling with the bubble (CB) modes are labeled by the plus ('$+$') and cross ('$\times$') symbols, respectively. 
The solid (red color) and dashed (blue color) lines show the upper and lower boundaries to star-forming galaxies from \citet{Kewley2001} and \citet{Kauffmann2003}, respectively, whereas those above the solid red line are classed as AGN. The plotted observations are from \citet{Richard2011}, \citet{Masters2014}, \citet{Berg2016}, and \citet{Senchyna2017}.
\label{fig:bpt:diag:radi}
}
\end{figure*}

\begin{figure*}
\begin{center}
\includegraphics[width=0.8\textwidth, trim = 0 0 0 0, clip, angle=0,origin=rb]{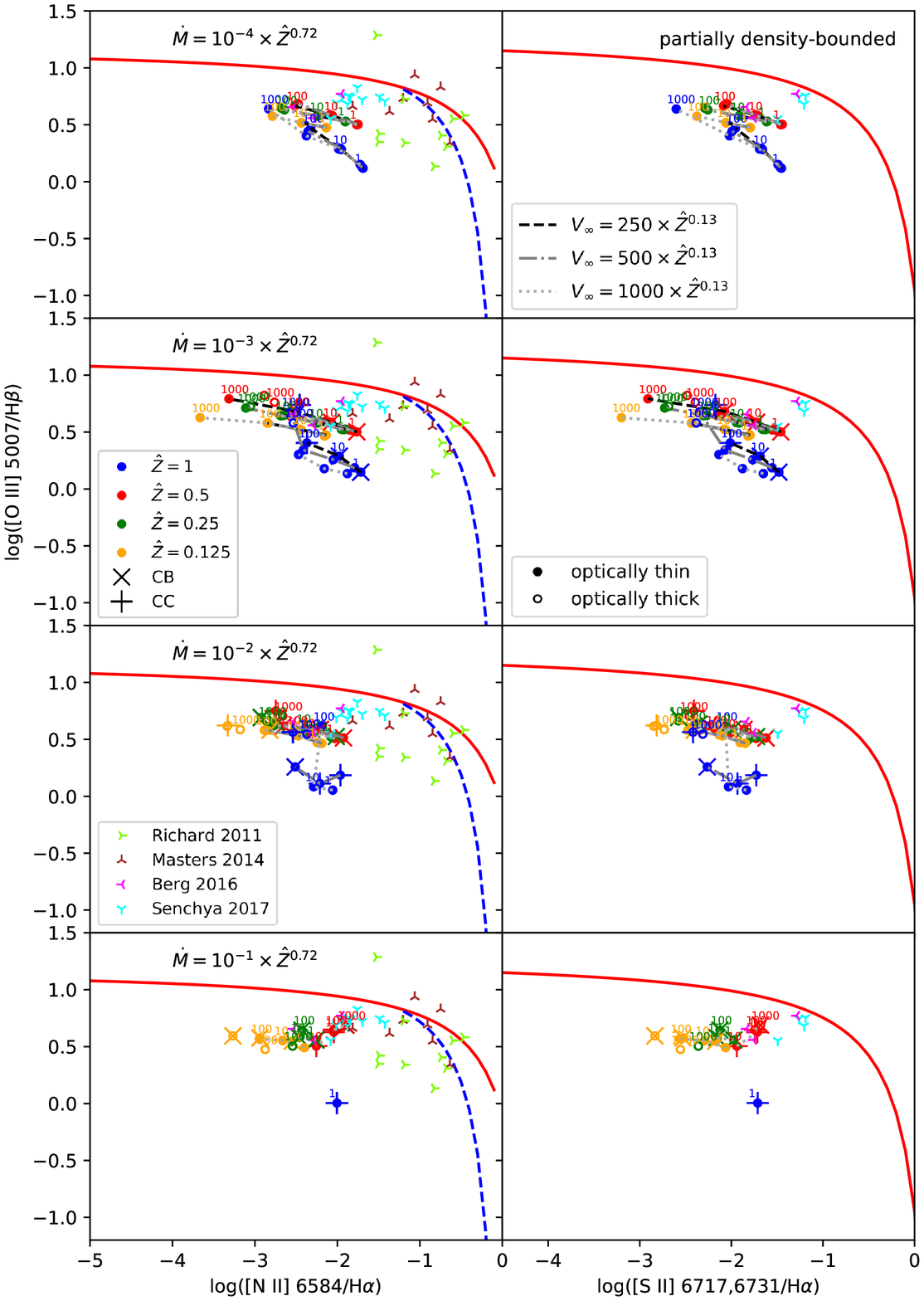}
\end{center}
\caption{The same as Figure~\ref{fig:bpt:diag:radi}, but for the partially density-bounded models.
\label{fig:bpt:diag:pden} 
}
\end{figure*}

\subsection{Diagnostic Diagrams}
\label{cooling:opt:diagnostics}

Here, we explore the emission-line parameter space generated by our CPI models in a few diagnostic diagrams, including the locus for catastrophic cooling models.  We use our total luminosities derived from \cloudy emissivities to produce diagnostic diagrams for different metallicities, wind terminal velocities, ambient densities, and mass-loss rates.

Figure~\ref{fig:bpt:diag:radi} shows ``BPT diagrams''
\citep{Baldwin1981} based on  \foiii $\lambda$5007/\hb
versus \fsii $\lambda\lambda$6717,6731/\ha and \fnii
$\lambda$6584/\ha\ for our fully radiation-bounded models with
optically thin and thick shells plotted by filled and empty circles,
respectively.
As noted above, given the uncertainties when the transition between
optically thin and thick conditions takes place in the shell, the line
emissivities for optically thick models are somewhat uncertain.
These diagrams are widely used to distinguish between active
galactic nuclei (AGN) and starburst galaxies
\citep[e.g.,][]{Veilleux1987,Osterbrock1992,Kewley2001,Kewley2006,Kewley2013,Groves2004,Groves2006}. 
Similarly, Figure~\ref{fig:bpt:diag:pden} shows
the same models as in Figure~\ref{fig:bpt:diag:radi}
but with the emission-line luminosities calculated as density-bounded
at the shell outer radius transverse to the line of sight, as
described above.

In general, we see that both sets of models in
Figures~\ref{fig:bpt:diag:radi} and \ref{fig:bpt:diag:pden} show line
ratios typical of photoionized \hii\ regions.
This applies to both adiabatic and strongly cooling conditions.
The models are highly
excited, although slightly below the starburst boundaries 
defined by \citet{Kewley2001} and \citet{Kauffmann2003}. 
This is likely caused by the fact that the ionization parameter is diluted by
the outflows in our models, which displace the gas to larger radii at
the shells.  Higher density models also have higher ionization
parameters, as evidenced by their low \fnii\ and \fsii\ emission.
As expected, at higher metallicity, \foiii/\hb\
decreases due to stronger cooling.
The partially density-bounded models in
Figure~\ref{fig:bpt:diag:pden} are similar to the fully
radiation-bounded ones in Figure~\ref{fig:bpt:diag:radi}, but lacking
the strongest \fnii\ and \fsii\ emission, which ordinarily dominates
in the outer regions.
Since the wind itself is more highly ionized and low density, the
contributions from kinematic heating are not apparent in the species shown by the BPT diagrams.

\begin{figure*}
\begin{center}
\includegraphics[width=0.75\textwidth, trim = 0 0 0 0, clip, angle=0,origin=rb]{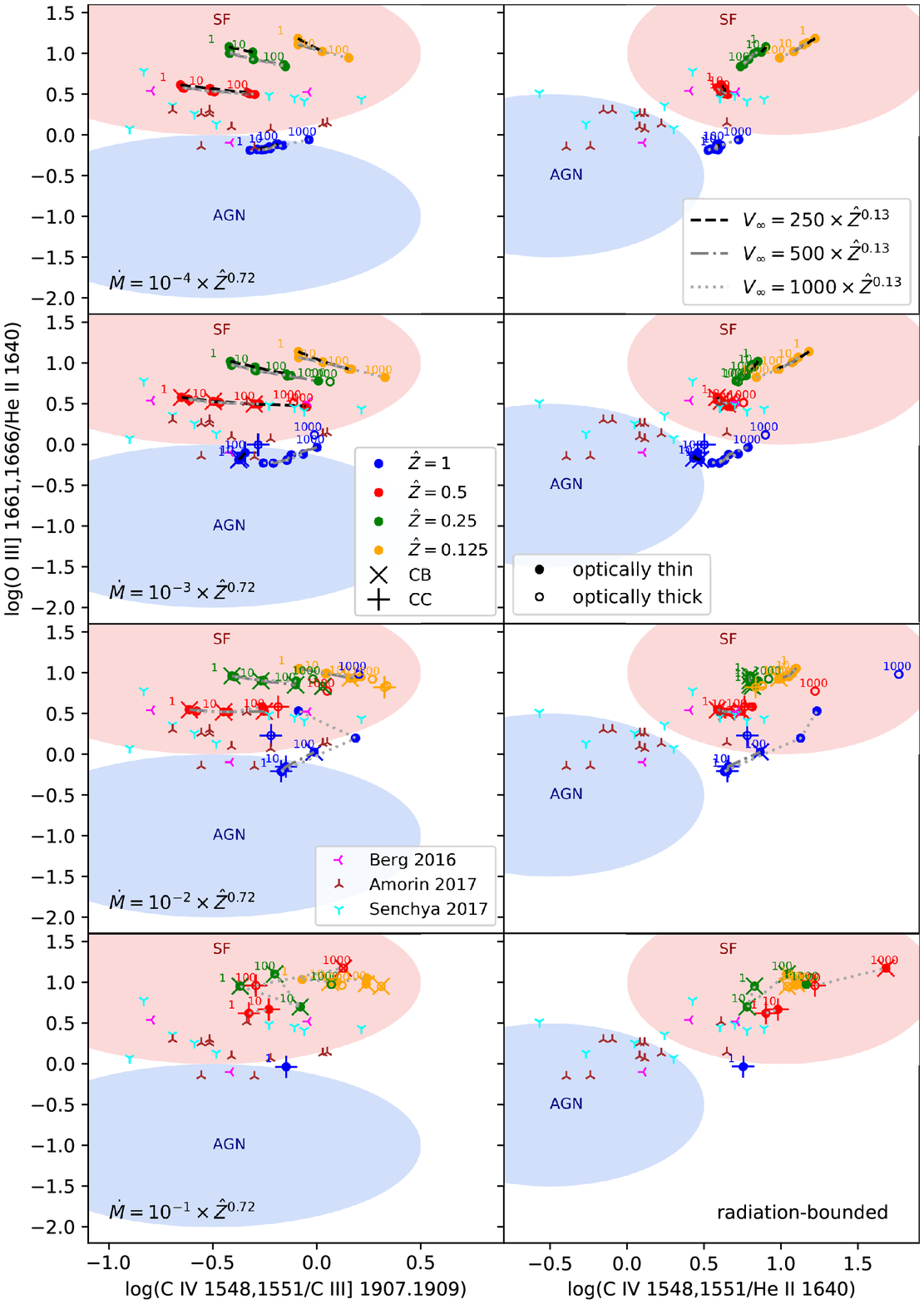}
\end{center}
\caption{The UV diagnostic diagrams plotted \ionf{O}{iii}] $\lambda\lambda$1661,1666/\heii $\lambda$1640 versus \ionic{C}{iv} $\lambda\lambda$1548,1551/\ionf{C}{iii}] $\lambda\lambda$1907,1909 (left) and  \ionf{O}{iii}] $\lambda\lambda$1661,1666/\heii $\lambda$1640 versus \ionic{C}{iv} $\lambda\lambda$1548,1551/\heii $\lambda$1640 (right panels).
Symbols and line types are as in Figure~\ref{fig:bpt:diag:radi}. 
The red and blue color filled areas represent the approximate regions associated with star-forming (SF) and active (AGN) galaxies from \citet{Gutkin2016} and \citet{Feltre2016}, respectively. 
The plotted observations are from \citet{Berg2016}, \citet{Amorin2017}, and \citet{Senchyna2017}.
\label{fig:uv:diag:radi}
}
\end{figure*}

\begin{figure*}
\begin{center}
\includegraphics[width=0.75\textwidth, trim = 0 0 0 0, clip, angle=0,origin=rb]{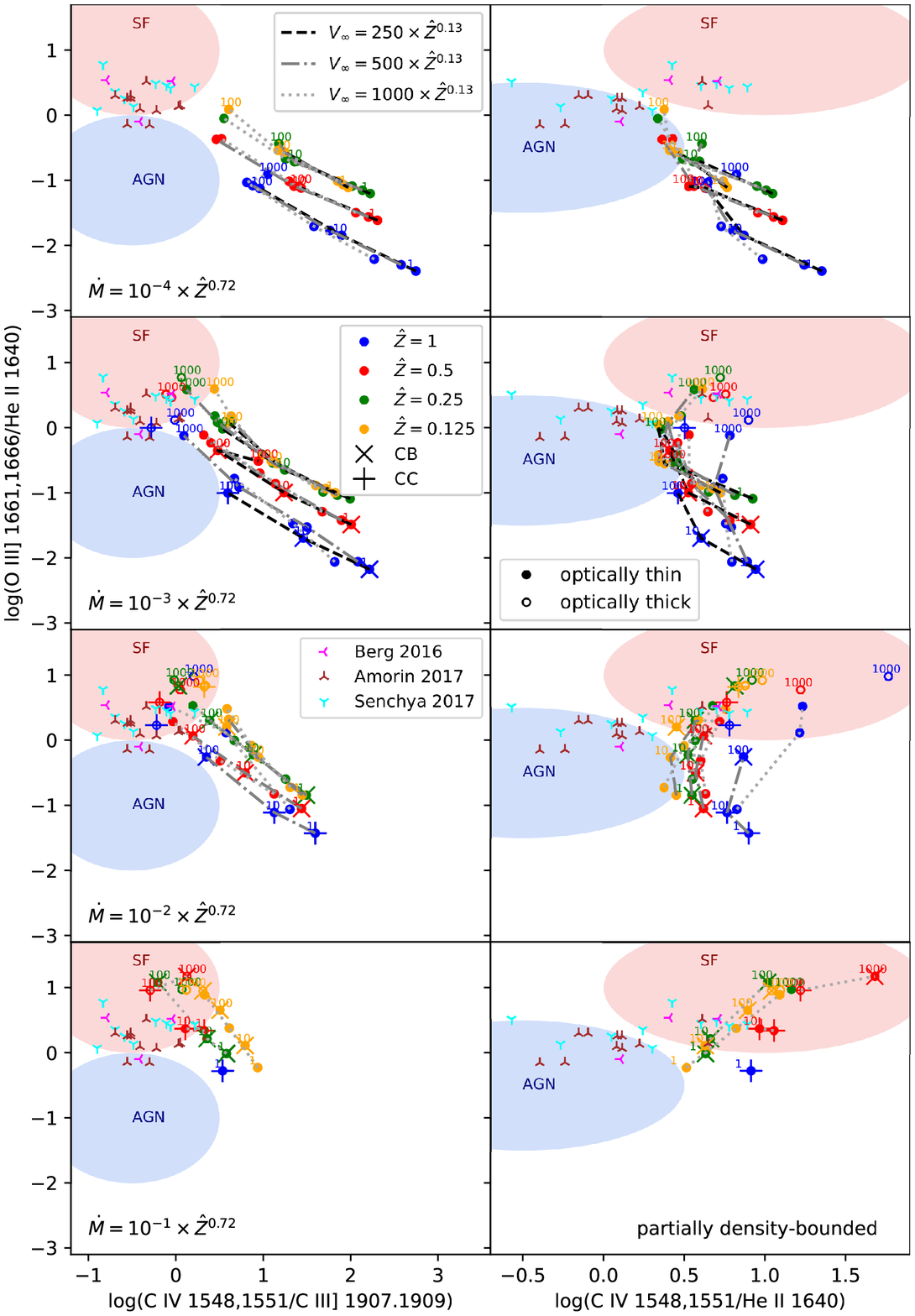}
\end{center}
\caption{The same as Figure~\ref{fig:uv:diag:radi}, but for the partially density-bounded models.
\label{fig:uv:diag:pden} 
}
\end{figure*}

\begin{figure}
\begin{center}
\includegraphics[width=0.42\textwidth, trim = 0 0 0 0, clip, angle=0,origin=rb]{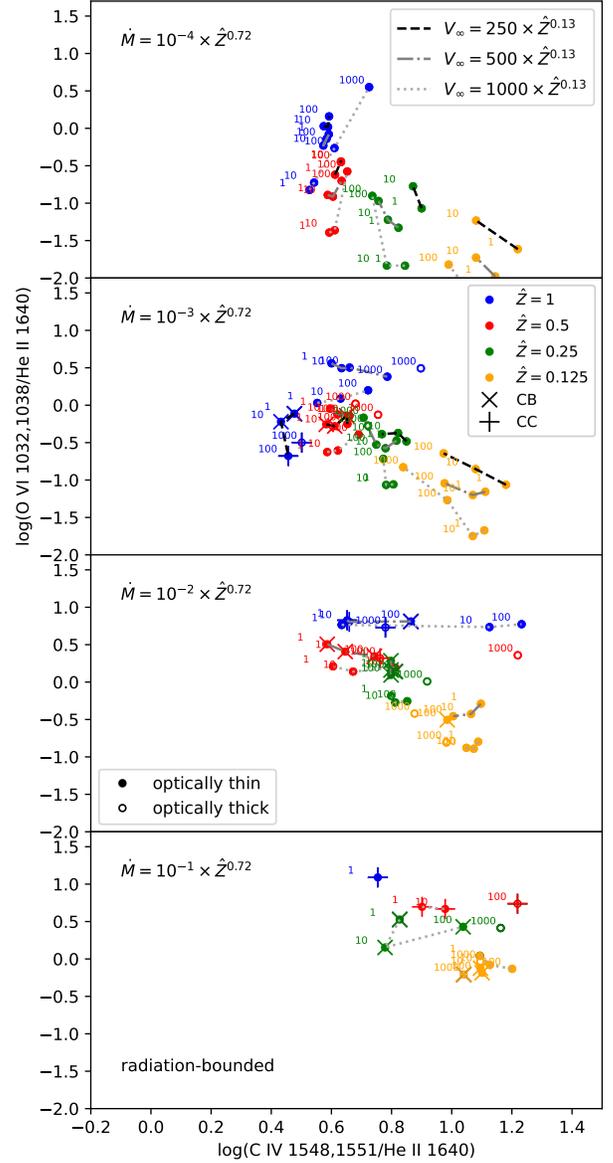}
\end{center}
\caption{
The UV diagnostic diagrams plotted \ovi $\lambda\lambda$1032,1038/\heii $\lambda$1640 versus \ionic{C}{iv} $\lambda\lambda$1548,1551/\heii $\lambda$1640 for the radiation-bounded models. 
Symbols and line types are as in Figure~\ref{fig:bpt:diag:radi}.
\label{fig:uv3:diag:radi}
}
\end{figure}

\begin{figure}
\begin{center}
\includegraphics[width=0.42\textwidth, trim = 0 0 0 0, clip, angle=0,origin=rb]{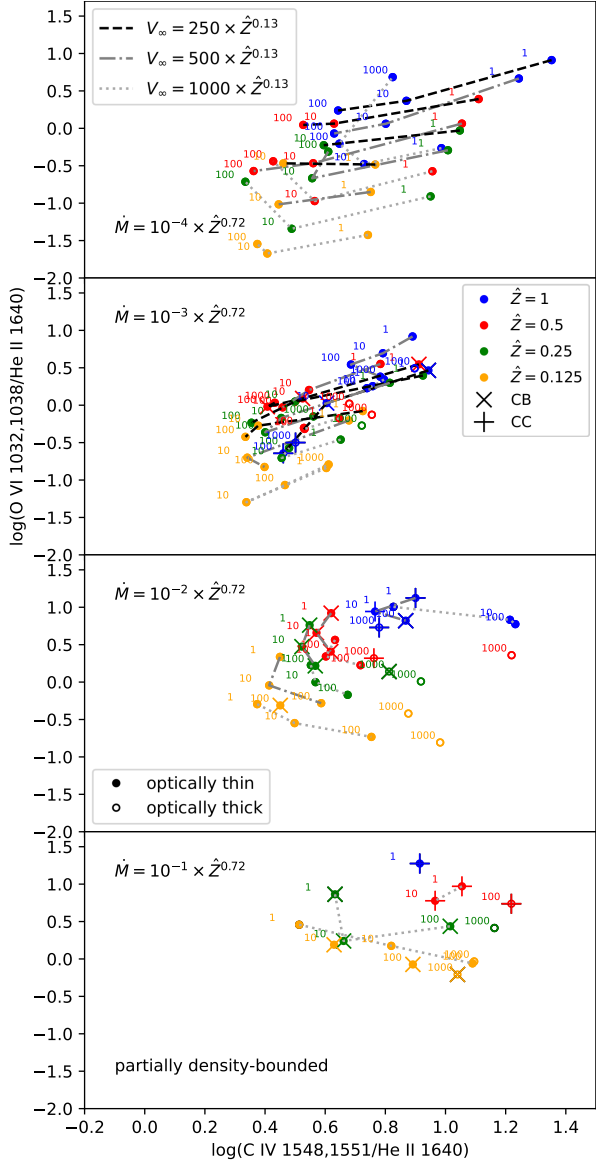}
\end{center}
\caption{The same as Figure~\ref{fig:uv3:diag:radi}, but for the partially density-bounded models.
\label{fig:uv3:diag:pden} 
}
\end{figure}

Similar to the optical BPT diagrams, UV diagnostics diagrams based on ultraviolet emission lines such as \civ $\lambda\lambda$1549,1551, \heii $\lambda$1640, \ionf{O}{iii}] $\lambda\lambda$1661,1666,   \ionf{C}{iii}] $\lambda\lambda$1907,1909, \heii $\lambda$1640 have been employed to distinguish between AGN and starburst galaxies \citep{Allen1998,Feltre2016,Gutkin2016,Hirschmann2019}, and radiative shock regions \citep{Allen1998,McDonald2015,Hirschmann2019}. We select sets of those UV diagnostic line ratios that are typically reported in star-forming regions to make UV diagnostic diagrams. Fully radiation-bounded models are presented in Figures~\ref{fig:uv:diag:radi} and \ref{fig:uv3:diag:radi}; and models that are density-bounded transverse to the line of sight in Figures~\ref{fig:uv:diag:pden} and \ref{fig:uv3:diag:pden}. 

Figure~\ref{fig:uv:diag:radi} presents the UV diagnostic diagrams for
\ionf{O}{iii}] $\lambda\lambda$1661,1666/\heii $\lambda$1640 versus
\ionic{C}{iv} $\lambda\lambda$1548,1551/\ionf{C}{iii}]
$\lambda\lambda$1907,1909 and \ionic{C}{iv}
$\lambda\lambda$1548,1551/\heii $\lambda$1640, for our fully
radiation-bounded models.
In general, these line ratios increase for decreasing metallicity,
due to the increasing nebular electron temperature in metal-poor
systems.  However, at $Z/$Z$_{\odot}=1$, strong adiabatic feedback
generates hot bubbles with strong \ionic{C}{iv} emission in the
interface with the dense shell, displacing some models for
$V_{\infty}=1000$ km\,s$^{-1}$ to higher \ionic{C}{iv} ratios
relative to other models.  For all other models, the wind heating is
unimportant for these emission-line ratios.  This highlights the
weakness of stellar winds at low metallicity.

These effects are more pronounced in Figure~\ref{fig:uv:diag:pden}, which is
the same as Figure~\ref{fig:uv:diag:radi}, but for models that are
partially density-bounded.  Since these models
are weighted toward emission in and near the shell, the \ionic{C}{iv}
emissivity is enhanced in models with significant adiabatic feedback,
such as those at lower density and higher velocity.  On the other hand,
\ionf{O}{iii}] is more sensitive to the overall electron temperature, which
decreases with increasing  the ambient density.

The highly ionized doublet \ovi $\lambda\lambda$1032,1038
originates from shocks and hydrodynamic collisional ionization, rather than
photoionization.  \citet{Gray2019} suggested that it could be a useful
diagnostic of catastrophic cooling flows.  In Figures~\ref{fig:uv3:diag:radi}
and \ref{fig:uv3:diag:pden},  we examine 
\ovi$\lambda\lambda$1032,1038/\heii $\lambda$1640 versus
\civ $\lambda\lambda$1549,1551/\heii $\lambda$1640,
for fully radiation-bounded, and partially density-bounded transverse to the
line of sight, respectively.  In our models, the \ovi\ emission is
produced primarily in the central region of the free-expanding wind,
nearest to the SSC; and in the dense shell interfaces to the hot
bubble region (see Figure~\ref{fig:CIE:emissivity}, top panels).

Figures~\ref{fig:uv3:diag:radi} and \ref{fig:uv3:diag:pden} show that
denser winds increase the \ovi\ emission through strong cooling, which correlates strongly
with  $\dot{M}_{\rm sc}$, as well as showing a slight anticorrelation with $V_{\infty}$.
These effects are enhanced in the partially density-bounded plots in Figure~\ref{fig:uv3:diag:pden}.
This is consistent with the suggestion by Paper~I that \ovi\ is
associated with catastrophic cooling, but we also see that adiabatic
models also show significant \ovi.  Thus, while \ovi\ is indeed
enhanced in most catastrophic cooling models, it is difficult to use
as a definitive diagnostic of these conditions without other
constraints on the parameter space.  Indeed, for the most strongly
cooling models, in which the hot bubble region is eliminated, the
temperature is too low to generate \ovi\ at all.

We caution that, as demonstrated by \citet{Gray2019}, the \ionic{O}{vi} emission line
can be strong in the outflow due to non-equilibrium ionization
states. Moreover, Paper~I shows that NEI conditions are
expected to contribute to more highly ionized states than in CIE, 
especially in strongly cooling flows.  
This is also seen, for example, in supernova remnants
\citep[e.g.][]{Patnaude2009,Patnaude2010, Zhang2019}.

In Figure~\ref{fig:bpt:diag:radi}--\ref{fig:uv3:diag:pden}, we also see that optically thick models typically appear
in denser ambient media. 
Figure~\ref{fig:bpt:diag:radi} shows that the ambient density of $n_{\rm amb}=10^3$\,cm$^{-3}$ results in weaker singly-ionized emission lines \fnii and \fsii. 
As seen in Figure~\ref{fig:uv:diag:radi}, an increase in the
metallicity ($Z/$Z$_{\odot}$) mostly tends to decrease the UV line ratios
\ionf{O}{iii}]/\heii\ and \ionic{C}{iv}/\heii, 
since the nebular temperature decreases.  
It is also not clear that enhanced \civ\ emission is strongly
associated with catastrophic cooling models as suggested in Paper~I.
However, future work with NEI models are needed to clarify this.

As noted above, our \cloudy\ models include dust grains with $M_{\rm d}/M_{\rm Z}=0.2$ found by \citet{DeVis2019} 
for evolved galaxies. 
Dust grains could also form in shocked ejecta from
SN explosions \citep[e.g.,][]{Nozawa2010} in starburst-driven superwinds.
On the other hand, the dust-to-metal ratio could be lower for unevolved galaxies. 
Although the optical diagnostic line ratios plotted in the BPT
diagrams are not greatly affected by the presence of dust grains,
they can have significant changes on the UV line ratios, since
dust grains slightly decrease radiation fields. 
Moreover, metal depletion onto dust grains reduces the ratios of 
\civ $\lambda\lambda$1549,1551/\heii $\lambda$1640 and
\ovi$\lambda\lambda$1032,1038/\heii $\lambda$1640.  
Thus dust grains and depletion onto them are additional sources of uncertainties 
in these line ratios, as well as in NEI cooling functions \citep[e.g.,][]{Richings2014}.


\section{Comparison with Observations}
\label{cooling:observations}

We compare our results 
with optical and UV observations of nearby and distant starburst galaxies, which were found to have 
detected observations of \heii $\lambda$1640, \civ $\lambda$1550, \ionf{O}{iii}]
$\lambda$1664, and \ionf{C}{iii}] $\lambda$1908.
The observations of low metallicity compact dwarf galaxies studied by \citet{Berg2016} 
have mean oxygen abundance of $12+\log($O/H$)\approx 7.6$ ($\hat{Z} \approx 0.125$), 
and electron temperatures $T_{\rm e}($[\ionf{N}{ii}]$) \approx 15000$\,K. 
Nearby starburst dwarf galaxies studied by
\citet{Senchyna2017} have similar parameters, but with $12+\log
($O/H$)\lesssim 8.3$ ($\hat{Z} \lesssim 0.6$).
More distant starburst galaxies have been extensively analyzed using optical nebular lines, and
some have been recently studied
through rest-frame UV spectroscopic observations 
\citep{Stark2014,Patricio2016,Steidel2016,Amorin2017,Vanzella2016,Saxena2020}. 
In particular, \citet{Amorin2017} considered 10 starburst galaxies at redshift $2.4< z< 3.5$ with 
mean metallicity of $12+\log($O/H$) \approx 7.6$, for which they measured  Ly$\alpha$ $\lambda$1216, 
\civ $\lambda$1550, \heii $\lambda$1640, \ionf{O}{iii}] $\lambda$1664, and \ionf{C}{iii}] $\lambda$1908. 
These objects have similar properties and line ratios as the samples
of \citet{Berg2016} and \citet{Senchyna2017}, and they are also
included in Figures~\ref{fig:uv:diag:radi} and \ref{fig:uv:diag:pden}.
Their properties are largely the same as those of the local sample
for the considered line ratios.
The galaxy samples occupy the parameter space used for our models,
as shown in Figures~\ref{fig:bpt:diag:radi} to \ref{fig:uv:diag:pden}.

It can be seen
in Figures~\ref{fig:bpt:diag:radi} and \ref{fig:bpt:diag:pden}
that, for the appropriate metallicity within
$Z/$Z$_{\odot} = 0.125$ and $0.5$, the data agree more with our fully
radiation-bounded models than with the partially density-bounded models.
This implies that the conditions in these objects do not require
significant density bounding for the integrated nebular emission.
However, since these apertures are on the order of a few kpc, we note that
the observed singly-ionized emission lines
could be due to the diffuse interstellar medium, and so some density
bounding is not ruled out.
These samples have similarities to the Green Pea galaxies, which tend
to have higher ionization parameters, with weaker emission from the low-ionization
species in Figures~\ref{fig:bpt:diag:radi} and \ref{fig:bpt:diag:pden}.

Figures~\ref{fig:uv:diag:radi}--\ref{fig:uv:diag:pden} also show the
UV observational data from \citet{Berg2016} and \citet{Senchyna2017}
plotted on the UV diagnostic diagrams. 
We again see reasonably good correspondence between the data and models for
$Z/$Z$_{\odot} = 0.5$ and 0.25.
In Figure~\ref{fig:uv:diag:radi},
the models appear to slightly overpredict 
\ionf{O}{iii}]/\ionic{He}{ii} ratio.
For these radiation-bounded conditions, the discrepancy may be due
to the CIE approximation in our models.  As shown in Paper~I, NEI
conditions for such outflows generate slightly higher ionization
states, which would reduce
\ionf{O}{iii}]/\ionic{He}{ii} and  \ionf{C}{iii}]/\ionic{C}{iv}. 
Furthermore, Figure~\ref{fig:uv:diag:pden} also shows that only a small degree of
density bounding for the high-density models can alleviate the
discrepancy.  This density bounding 
could be an observational bias, as suggested earlier; or it could be
at least partially real, since similar galaxies have been found to be
Lyman-continuum emitters \citep[e.g.,][]{Izotov2016,Izotov2018}.

The \ionic{O}{vi} $\lambda\lambda$1032,1038 emission and absorption
lines are more readily associated with active galaxies and AGN
feedback, rather than starbursts 
\citep[e.g.,][]{Baldwin2003,Hainline2011,Tripp2008,Savage2014,Bielby2019}.  
We do not typically expect \ionic{O}{vi} emission in ordinary \ionic{H}{ii}
regions, since these highly ionized lines originate from a hot gas phase with $T \sim
3 \times 10^5$\,K. Generally in starbursts, this species is only seen in absorption,
associated with adiabatic outflows \citep[e.g.,][]{Heckman2001}.
But recently, \citet{Hayes2016} spatially resolved \ionic{O}{vi} $\lambda\lambda$1032,1038 emission 
in a nearby, intense starburst galaxy (SDSS
J115630.63+500822.1, hereafter J1156) through \textit{Hubble Space
  Telescope} (HST) imaging, having spectroscopically observed this
emission with the Far Ultraviolet Spectroscopic Explorer (FUSE). Moreover, they also detected a blueshifted \ionic{O}{vi} absorption line associated with an average outflow velocity of $380$\,km\,s$^{-1}$. 
From the \ionic{O}{vi} column density, they found an
\ionic{O}{vi} mass of $3 \times 10^4$ M$_{\odot}$.
The starburst properties of this galaxy are similar to those of our
comparison datasets, and it has metallicity $7.87 < 12+\log ($O/H$)< 8.34$ (i.e. $0.23 < \hat{Z} < 0.7$).

\citet{Chisholm2018} confirmed that photoionization models of galactic
outflows with \cloudy\ cannot reproduce the observed \ionic{O}{vi}
absorption in a high-redshift starburst, 
so they again point to the standard model that \ionic{O}{vi} originates
from a mass-loading interface between the photoionized region and
unobserved hot wind.
However, \citet{Gray2019} 
found that NEI models can produce 
a range of \ionic{O}{vi} column densities in dense outflows themselves,
confirming that
relatively high \ionic{O}{vi} emission 
can also be due to hydrodynamic heating.
Moreover, Paper~I also reproduced 
strong \ionic{O}{vi} and \ionic{C}{iv} emission from strongly cooling winds, so
they suggested that these lines could be used to trace catastrophically cooling superwinds in starburst regions. 
In Figure~\ref{fig:uv3:diag:radi}, we see that
the \ovi\ emission is indeed enhanced by increasing the mass-loss rate and
metallicity, and it is typically stronger in catastrophic cooling
models.  However, the \ovi\ emission plummets in the models at
$Z/$Z$_{\odot}=1$ and $\dot{M}_{\rm sc}$\,$=$\,$10^{-3}$\,M$_{\odot}/$yr, where strong radiative cooling  
occurs.  Moreover, \ovi\ is also seen in adiabatic
models, so additional constraints on the parameter space are necessary
when using this line to evaluate the presence of catastrophic cooling.

\section{Discussion}
\label{cooling:discussion}

The local extreme emission-line galaxies are regarded as analogs of higher redshift starbursts. Various samples of such objects have been reported, and they have similar properties and emission-line ratios as the comparison samples above from \citet{Berg2016} and \citet{Senchyna2017}.  For example, the recent deep VANDELS spectroscopic survey of \ionic{He}{ii} emitting galaxies by \citet{Saxena2020} identified 8 objects with high enough SNR at $z \sim 2.5$--3.5 to detect \ionic{He}{ii} $\lambda$1640, \ionf{O}{iii}] $\lambda\lambda$1661,1666 lines, and \ionf{C}{iii}] $\lambda$1909 in emission. The metallicity of these \ionic{He}{ii} emitters are estimated to be in the range $Z/$Z$_{\odot}=0.125$ to $0.25$, similar to the low-redshift samples. Other UV observations of intermediate redshift sources have similar physical conditions. Deep spectroscopic observations of low mass, gravitationally lensed galaxies at $z \sim 1.5$--3 conducted by \citet{Stark2014} identified moderately metal-poor gaseous environments with $ \sim 0.2$--0.5 Z$_{\odot}$, and MUSE integral field spectroscopy of a young lensed galaxy at $z = 3.5$ (SMACSJ2031.8-4036) by \citet{Patricio2016} revealed similar physical conditions ($T_{\rm e} \sim 15600$\,K) and metallicity ($\lesssim 0.16$ Z$_{\odot}$).

These highly excited starbursts appear to be linked to Lyman continuum (LyC) emission.  The extreme Green Pea galaxies are to date the only class of objects known to be confirmed LyC emitters \citep[e.g.,][]{Izotov2016, Izotov2018}, and they are characterized by their extreme nebular excitation \citep{Cardamone2009}. At intermediate redshift, \citet{deBarros2016} investigated the UV spectrum of a LyC emitter candidate at $z=3.2$, called \textit{Ion2}, which has  $12+\log($O/H$) \approx 8.1$ ($\hat{Z} \approx 0.4$), and strong Ly$\alpha$ \foiii $\lambda\lambda$4959,5007 and \ionf{C}{iii}] $\lambda\lambda$1907,1909. \textit{Ion2} exhibits an extremely high \foiii/\foii ratio $>10$, which is much higher than expected according to the SFR--$M_{\star}$--$Z$ dependence of \foiii/\foii ratios derived by \cite{Nakajima2014}. \citet{deBarros2016} suggest that this could be due to a density-bounded scenario. Indeed, primeval galaxies at the time of cosmic reionization ($z > 7$) also have these extreme emission line properties, and similarly are very compact, with high star formation rates and lower masses \citep[e.g.,][]{Grazian2015,Shibuya2015,Tasca2015}. \citet{Stark2015} analyzed the UV observations of a gravitationally lensed, Ly$\alpha$ emitter at $z = 7.045$ (A1703-zd6) and having $12+\log($O/H$) \approx 7.04$. They detected \ionic{C}{iv} $\lambda$1548 emission with ${\rm FWHM} \lesssim 125$\,km\,s$^{-1}$. 
Although this is similar to measurements in narrow-lined AGNs at lower redshift,
it could also originate from a starburst region powered by young, very hot, metal-poor stars.

Catastrophic cooling conditions have been suggested to be
present in some of these extreme starbursts
\citep[e.g.,][]{Silich2020,Jaskot2019,Oey2017} based on kinematic and
other evidence.  If this is indeed the case, then our results would imply that
mostly likely the heating efficiency in these systems is significantly reduced
relative to the fiducial scalings for stellar wind velocities
(Section~\ref{cooling:hydrodynamics:results}).  It is also possible
that mass-loading plays a role, but the degree of mass
loading required to induce catastrophic cooling is high.  Since these
objects tend to be metal-poor, it is unlikely that metallicity plays a
significant role in promoting cooling.
If this is indeed the case, then our results in
Section~\ref{cooling:hydrodynamics:results} would imply that either
(1) the heating efficiency in these systems is low; (2) mass-loading
is high; or (3) the winds are suppressed within the injection radius
$R_{\rm sc}$.  The first option would imply that the effective wind
velocity is reduced by at least a factor of 2--4, since stellar winds
are on the order of 1000--2000 km\,s$^{-1}$, even at low metallicity.
This would correspond to a decrease in kinetic heating efficiency of 0.25 to
0.06, and could be caused by a variety of factors like entrainment of
clouds, turbulence, and magnetic fields.  The second option would
similarly require an increase in effective $\dot{M}_{\rm sc}$ by an
order of magnitude, which could be accomplished by mass-loading.  The
third option would require both weak velocities and high $\dot{M}_{\rm
  sc}$ in order to suppress the wind within $R_{\rm sc}$.  This has
been suggested to be the case in NGC 5253-D1 by \citet{Silich2020}.  

Thus, our modeled emission-line spectra and outflow parameter space
may be relevant to LyC emitters and primordial galaxies.  
Our comparison with the local extreme starbursts shows that our models for
$Z/Z_\odot = 0.5$ and 0.25 are reasonably good at matching the
observed emission-line data for \ionic{C}{iv}, \ionf{C}{iii}], 
  \ionf{O}{iii}], and \ionic{He}{ii} when invoking a slight density bounding 
  at high ambient densities, 
  on the order of 100 cm$^{-3}$.  This
density bounding  would be expected for LyC emitters, although it 
could also simply be an observational bias favoring the
brightest, swept-up gas.  Or, as discussed in
Section~\ref{cooling:observations}, the objects are fully radiation
bounded but with NEI conditions in the outflow that cause the modest
variation from our modeled emission-line ratios.  Both scenarios imply
a significant contribution from kinematic heating to generate the
observed high ionization states.
Indeed, catastrophic cooling has been suggested to be linked to
LyC-leaking galaxies \citep{Jaskot2017, Jaskot2019}. 

As suggested by Paper~I, catastrophic cooling models as a group tend
to show higher emission in \ionic{C}{iv} and \ionic{O}{vi}, especially
for the CB models.  But for stronger cooling in the CC models, the
temperature is too low to support strong emission in these ions.  We
also see that adiabatic models also produce \ionic{C}{iv} and
\ionic{O}{vi}, complicating their use as diagnostics of catastrophic
cooling, although additional work is needed to evaluate this.
Since our models are still limited in parameter space, our
comparisons with observations are not conclusive, but illustrate the
feasibility of the catastrophic cooling and density bounding as
important factors in observed line emissivities.
Further study of known LyC emitters with resolved nebular data, such
as Haro 11 \citep[e.g.,][]{Micheva2020,Keenan2017} is also needed to
develop robust nebular diagnostics of these conditions.

\section{Conclusions}
\label{cooling:conclusion}

In this paper, we employ the non-equilibrium atomic chemistry and
cooling package \maihem \citep{Gray2015,Gray2016,Gray2019} to
investigate how superwinds driven by super star clusters
are subject to radiative cooling. 
We assume a SSC with mass $2.05\times10^6$\,M$_{\odot}$, radius of 1
pc, and age of 1\,Myr, and we model a range of ambient density  ($n_{\rm
  amb}$\,$=$\,$1,\ldots,10^3$\,cm$^{-3}$).
At solar metallicity, the superwind is parameterized by mass-loss
$\dot{M}_{\rm sc}$\,$=$\,$10^{-1},\ldots,10^{-4}$\,M$_{\odot}/$yr, and
wind terminal velocity  $V_{\infty}$\,$=$\,$250,500,1000$\,km\,s$^{-1}$).
At lower metallicity ($Z/$Z$_{\odot}=0.5, 0.25,0.125$), $\dot{M}_{\rm
  sc}$ and $V_{\infty}$ are scaled by the metallicity dependencies, i.e.
$\dot{M}_{\rm sc} \varpropto Z^{0.72}$ and $V_{\infty} \varpropto Z^{0.13}$. 
The superwind model is implemented following \citet{Gray2019a},
according to the radiative solution presented by \citet{Silich2004}.

Our resulting grid of superwind models 
demonstrates where strongly radiative cooling occurs in this parameter space.
The superwind structures produced by our hydrodynamic simulations are
classified 
according to their departure from the expected adiabatic temperature,
as well as the formation of the characteristic bubble within the
simulated time frame.  
Our models are parameterized in terms of $\dot{M}_{\rm sc}$
and $V_{\infty}$, which account for
mass-loading and/or heating efficiency effects.

We find that decreasing $V_{\infty}$, or equivalently, the kinetic
heating efficiency, significantly enhances
radiative cooling effects for a given fixed $\dot{M}_{\rm sc}$ and $n_{\rm
  amb}$ (Figure~\ref{fig:cooling:temp:fact}).
Similarly, high mass-loss rates ($\dot{M}_{\rm sc}$), or equivalently,
high mass loading, can lead to catastrophic cooling in 
slower stellar winds (lower $V_{\infty}$).
Where both of these conditions apply,
radiative cooling strongly dominates and prevents any dynamically
significant thermal energy contribution, resulting in either
momentum-conserving outflows, or 
winds failing to launch from the cluster injection radius $R_{\rm sc}$. 
Increasing $Z$ also promotes cooling, but it is not as strong an effect
as the wind parameters. 

We also find that the presence of the distinctive, hot bubble
morphology is not always a reliable indicator of adiabatic status for
the outflow.  The hot bubble accompanies both
adiabatic outflows and ones with strong cooling flows.  Therefore, the
existence of a hot superbubble is not necessarily indicative of adiabatic
feedback.  Similarly, some adiabatic models take time to develop the bubble
morphology. 
Thus, for young systems with ages $\lesssim 1$ Myr, the
lack of a hot, X-ray bubble is not necessarily an indicator that the
system is not adiabatic. 

We calculate
line emissivities for the density and temperature profiles produced by
our superwind models using the photoionization code \cloudy
\citep{Ferland1998,Ferland2013,Ferland2017}. 
The line luminosities are computed from the volume emissivities for
fully radiation-bounded models, and models that are density-bounded
transverse to the line of sight.
We construct a few 
optical and UV diagnostic diagrams from the predicted 
line luminosities (Figures~\ref{fig:bpt:diag:radi}
--\ref{fig:uv3:diag:pden}) to compare with observed
emission lines from extreme starbursts. 
As noted above, dust is also a factor in evaluating the UV line ratios. 

As suggested by Paper~I,  \ionic{O}{vi} and \ionic{C}{iv} are stronger
in modest catastrophic cooling flows, especially the former.  For
extreme cooling, the temperature is too low to generate strong
emission in these ions.  Since
adiabatic models also produce these emission lines, their use
as diagnostics of catastrophic cooling is complicated, and
requires careful analysis of the parameter space.  Further study is
needed to identify unambiguous diagnostics.

Our optical and UV diagnostic line ratios predicted by 
radiation-bounded models are in general agreement with observations of
nearby and distant star-forming galaxies having similar physical
properties (see Figures~\ref{fig:bpt:diag:radi}
--\ref{fig:uv3:diag:pden}).
The modest discrepancies observed in our
UV diagnostic diagrams in \ionic{C}{iv}, \ionf{C}{iii}], \ionf{O}{iii}], and  \ionic{He}{ii} are
consistent with minor density bounding, implying LyC escape, as has
been suggested for similar galaxy populations.  Or, the observed
density bounding could be an observational bias toward the dense,
piled-up shells.  NEI conditions may also be responsible.  However,
all of these interpretations imply significant emission from
kinetically heated thermal components in our modeled outflows.

We caution that our 1-D hydrodynamic simulations do not
reproduce thermal and dynamical instabilities that create clumping in the
free-expanding superwind and shell.
This would significantly affect the outflow morphology and evolution,
and by extension, the expected emission-line spectra.
Instability-induced gas clumping within the outflow may be important
\citep[e.g.,][]{Jaskot2019}, and
needs to be investigated through 2-D hydrodynamic simulations. 
Other important parameters remain to be explored, including the effect of 
the SSC size, ambient density gradient, system age, and non-equilibrium ionization.
Our photoionization
calculations are implemented using the typical dust-to-metal mass
ratio associated with evolved galaxies, whose metallicity dependence
is not well understood.
Future computations are
required to investigate these effects in order to 
better understand the occurrence and properties of catastrophic cooling in starburst
regions. 

\begin{acknowledgments}

We are grateful to Sergiy Silich for useful discussions and comments
on the manuscript.  We thank
the referee for comprehensive comments and suggestions. 
M.S.O. acknowledges NASA grants HST-GO-14080.002-A and HST-GO-15088.001-A.

The hydrodynamics code \flash used in this work was developed in part by the DOE 
NNSA ASC- and DOE Office of Science ASCR-supported Flash Center for Computational
Science at the University of Chicago.
Analysis and visualization of the \flash simulation data were 
performed using the yt analysis package \citep{Turk2011}.

\end{acknowledgments}



\software{\flash \citep{Fryxell2000}, yt \citep{Turk2011}, \cloudy \citep{Ferland2017}, Starburst99 \citep{Leitherer2014}, NumPy \citep{Harris2020}, SciPy \citep{Virtanen2020}, Matplotlib \citep{Hunter2007}, HDF5 \citep{Folk2011}, \textit{hypre} \citep{Falgout2002}.}

\begin{appendix}

\section{2D Surface Brightness and Luminosity Calculations}
\label{appendix:a}

For comparison of \cloudy results with observations, we need to generate the projected 2D surface brightness and
the total luminosity from the volume emissivity created by \cloudy for each emission line. We can then compare the observed spatially resolved flux with the projected 2D surface brightness, and the observed integrated flux with the total luminosity.

\begin{figure}
\begin{center}
\includegraphics[width=0.28\textwidth, trim = 40 0 80 0, clip, angle=270]{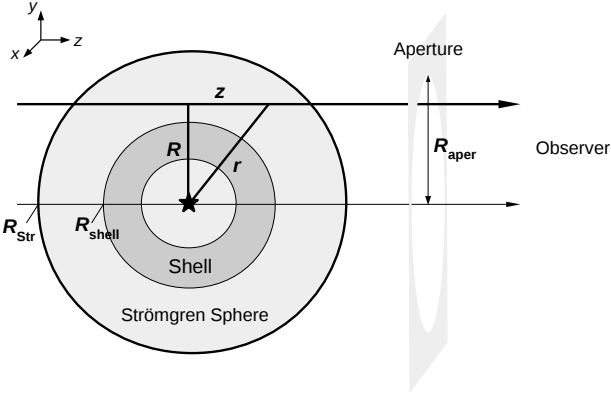}
\end{center}
\caption{Schematic view of the construction of the projected surface brightness $I_{\lambda} (R)$ as a function of projected radius $R$  according to the volume emissivity $\epsilon_{\lambda} (r)$  as a function of radius $r$. The line of sight is along the $z$-axis, and the object is projected onto the $x$-$y$ plane. The object is a spherical symmetric \ionic{H}{ii} region having Str\"{o}mgren radius $R_{\rm Str}$ with a spherical symmetric shell having maximum radius of $R_{\rm shell}$.  The observer measures the total luminosity over a circular aperture with radius of $R_{\rm aper}$ partially (or entirely) covering the Str\"{o}mgren sphere (or the shell).
\label{fig:surface:brightness}
}
\end{figure}

We set the program \cloudy to generate the volume emissivity $\epsilon_{\lambda} (r)$ at a given radius $r$ for each emission line at wavelength $\lambda$. Figure~\ref{fig:surface:brightness} shows a schematic of the transformation from the volume emissivity $\epsilon_{\lambda} (r)$ to the surface brightness $I_{\lambda} (R)$ projected onto the $x$-$y$ plane perpendicular to the line of sight along the $z$-axis. We assume that the \ionic{H}{ii} region is  optically thin and spherically symmetric. To calculate the surface brightness of a spherical symmetric object at each point on $x$ and $y$ projected plane, we employ the SciPy package to integrate the volume emissivity $\epsilon_{\lambda} (r)$ over the $z$-axis, $I_{\lambda} (R) = 2 \int^{z_{\rm max}}_{z=0}  \epsilon_{\lambda} (r) dz$, where $z=\sqrt{r^2 - R^2}$ and $dz=rdr /\sqrt{r^2 - R^2}$, so the projected surface brightness  at the projected radius from the center of the object is calculated by 
\citep[see e.g.][]{Sarazin1986,Cimatti2019}
\begin{equation}
I_{\lambda} (R) = 2  \int^{R_{\rm max}}_{r=R}  \frac{\epsilon_{\lambda} (r)}{ \sqrt{r^2 - R^2} }  r dr,
\label{eq:21}%
\end{equation}
where $r$ is the radial distance of the volume emissivity from the center of the object, $R$ is the projected radius of a point on the 2D projected plane from the center of the object, and $R_{\rm max}$ the maximum radius of the geometry where the integral is performed over  and the volume emissivity $\epsilon_{\lambda} (r)$ should be also available at the maximum radius ($r= R_{\rm max}$).

We note that Eq. (\ref{eq:21}) is an Abel integral, so it can be inverted and deprojected to the volume emissivity as a function of radius as follows 
\citep[see e.g.][]{Cavaliere1980,Longair2008}:
\begin{equation}
\epsilon_{\lambda} (r) = - \frac{1}{\pi r}  \frac{d}{dr} \int^{R_{\rm max}}_{R=r}  \frac{I_{\lambda} (R)}{ \sqrt{R^2 - r^2 } }  R dR.
\label{eq:22}%
\end{equation}

The total luminosity integrated over a circular aperture entirely or partially covering a spherical symmetric \ionic{H}{ii} region (see Figure~\ref{fig:surface:brightness}) with the surface brightness $I_{\lambda} (R)$ projected on the 2D plane is
\begin{equation}
L_{\lambda} =  \int^{2 \pi}_{\varphi=0} \int^{R_{\rm aper}}_{R=0}  I_{\lambda} (R)  R d R d\varphi,
\label{eq:23}%
\end{equation}
where $R_{\rm aper}$ is the radius of the circular aperture where the integral is performed over. Substituting Eq. (\ref{eq:21}) into the above integral (\ref{eq:23}), we restore Eq. (\ref{eq:20}). 

We use Eq. (\ref{eq:21}) to construct the 2D projected surface brightness that is comparable with the observed spatially resolved emission-line flux. Moreover, we utilize Eq. (\ref{eq:23}) to calculate the total luminosity, which can be compared with the observed integrated emission-line flux.  

To model the observed emission that is radiation bounded in the line
of sight but density bounded in  transverse directions, we set 
$R_{\rm aper}=R_{\rm shell}$ and $R_{\rm max}=R_{\rm Str}$, i.e. a partially density-bounded model, 
however, $R_{\rm aper}=R_{\rm max}=R_{\rm shell}$ for a fully density-bounded model,  
and $R_{\rm aper}=R_{\rm max}=R_{\rm Str}$ for fully radiation-bounded model, 
where $R_{\rm shell}$ is the outer radius of the dense shell and 
$R_{\rm Str}$ is the Str\"{o}mgren radius (see Figure~\ref{fig:surface:brightness}).

\section{Supplementary Material}
\label{appendix:b}

The compressed (tar.gz) files of the interactive figure (64 images) of Figure~\ref{fig:temp:dens:profiles}, the interactive animation (64 videos) of Figure~\ref{fig:temp:dens:profiles2}, and the interactive figure (192 images) of Figure~\ref{fig:CIE:emissivity} are available in the electronic edition of this article, and are archived on Zenodo (doi:\href{https://doi.org/10.5281/zenodo.4989577}{10.5281/zenodo.4989577}). These interactive figures and animation are also hosted on this URL.\footnote{\url{https://superwinds.astro.lsa.umich.edu/}}

The compressed (tar.gz) file containing the 6 machine-readable tables for the emission-line data (partially presented in Table~\ref{tab:cloudy:output}) is available in the electronic edition of this article. 
Each file is named as \verb|table_case_bound.dat| such as
\verb|table_CPI_radi.dat|, where \verb|case| is for the
ionization case (\verb|PI|:
purely photoionization and \verb|CPI|: photoionization and hydrodynamic collisional
ionization), and \verb|bound| for the
optical depth model (\verb|radi|: fully radiation-bounded, \verb|pden|: partially density-bounded, and \verb|dens|: fully density-bounded). Each file contains the following information:

\begin{table}
\begin{center}
\caption[]{Integrated luminosities on a logarithmic scale (unit in erg/s) from the different ionization models (see Appendix~\ref{appendix:b} for more information). 
\label{tab:cloudy:output}}
\scriptsize
\begin{tabular}{l|cc|cc|cc}
  \hline\hline\noalign{\smallskip}
\multicolumn{1}{c|}{} & \multicolumn{2}{c|}{radiation-bounded} & \multicolumn{2}{c|}{part.density-bound} & \multicolumn{2}{c}{density-bounded} \\
\multicolumn{1}{c|}{case:} & PI & CPI & PI & CPI & PI & CPI  \\
Emission Line &    &    &    &    &    &    \\
\noalign{\smallskip}
\tableline
\noalign{\smallskip}
Ly$\alpha$     $\lambda$1216 &  40.836 &  40.577 &  40.443 &  40.191 &  40.317 &  40.077 \\
H$\alpha$      $\lambda$6563 &  40.789 &  40.789 &  40.460 &  40.459 &  40.181 &  40.179 \\
H$\beta$      $\lambda$4861 &  40.350 &  40.351 &  40.021 &  40.021 &  39.743 &  39.742 \\
\ionic{He}{i}     $\lambda$5876 &  39.435 &  39.440 &  39.112 &  39.112 &  38.834 &  38.833 \\
\ionic{He}{i}     $\lambda$6678 &  38.868 &  38.873 &  38.544 &  38.544 &  38.266 &  38.264 \\
\ionic{He}{i}     $\lambda$7065 &  39.111 &  39.119 &  38.812 &  38.814 &  38.527 &  38.525 \\
\ionic{He}{ii}    $\lambda$1640 &  38.801 &  38.868 &  38.761 &  38.805 &  38.692 &  38.733 \\
\ionic{He}{ii}    $\lambda$4686 &  37.011 &  37.031 &  37.003 &  37.018 &  36.992 &  37.005 \\
\ionic{C}{ii}     $\lambda$1335 &  38.403 &  38.401 &  37.939 &  37.937 &  37.540 &  37.538 \\
\ionic{C}{ii}     $\lambda$2326 &  38.390 &  38.177 &  37.537 &  37.360 &  35.933 &  35.926 \\
\ionic{C}{iii}    $\lambda$977  &  38.569 &  38.571 &  38.342 &  38.341 &  38.110 &  38.108 \\
\ionf{C}{iii}]   $\lambda$1909 &  39.880 &  39.880 &  39.257 &  39.265 &  38.672 &  38.686 \\
\ionic{C}{iii}    $\lambda$1549 &  36.898 &  36.899 &  36.674 &  36.673 &  36.443 &  36.441 \\
\ionic{C}{iv}     $\lambda$1549 &  39.419 &  39.611 &  39.111 &  39.423 &  38.940 &  39.343 \\
{[\ionf{N}{i}]}      $\lambda$5200 &  36.941 &  36.092 &  36.019 &  35.172 &  31.009 &  31.001 \\
{[\ionf{N}{ii}]}     $\lambda$5755 &  37.478 &  37.051 &  36.590 &  36.207 &  34.732 &  34.724 \\
{[\ionf{N}{ii}]}     $\lambda$6548 &  38.579 &  38.179 &  37.693 &  37.336 &  35.991 &  35.985 \\
{[\ionf{N}{ii}]}     $\lambda$6583 &  39.049 &  38.649 &  38.163 &  37.806 &  36.461 &  36.455 \\
\ionic{N}{iii}    $\lambda$1750 &  39.342 &  39.344 &  38.674 &  38.683 &  37.972 &  37.983 \\
\ionic{N}{iii}    $\lambda$991 &  39.056 &  39.059 &  38.816 &  38.816 &  38.578 &  38.576 \\
\ionic{N}{iv}     $\lambda$1486 &  39.152 &  39.210 &  38.770 &  38.861 &  38.470 &  38.620 \\
\ionic{N}{v}      $\lambda$1240 &  38.050 &  39.040 &  38.050 &  39.040 &  38.050 &  39.040 \\
\ionic{O}{i}      $\lambda$1304 &  38.206 &  38.011 &  37.289 &  37.092 &  33.788 &  33.784 \\
{[\ionf{O}{i}]}      $\lambda$6300 &  37.610 &  36.234 &  36.688 &  35.323 &  31.560 &  31.552 \\
{[\ionf{O}{i}]}      $\lambda$6364 &  37.114 &  35.739 &  36.193 &  34.827 &  31.064 &  31.056 \\
{[\ionf{O}{ii}]}     $\lambda$3726 &  38.947 &  38.609 &  38.105 &  37.841 &  36.955 &  36.949 \\
{[\ionf{O}{ii}]}     $\lambda$3729 &  39.093 &  38.753 &  38.248 &  37.980 &  37.055 &  37.049 \\
{[\ionf{O}{ii}]}     $\lambda$7323 &  37.601 &  37.363 &  36.931 &  36.812 &  36.380 &  36.377 \\
{[\ionf{O}{ii}]}     $\lambda$7332 &  37.518 &  37.280 &  36.846 &  36.726 &  36.290 &  36.288 \\
{\ionf{O}{iii}]}    $\lambda$1661 &  38.783 &  38.800 &  38.238 &  38.275 &  37.791 &  37.865 \\
{\ionf{O}{iii}]}    $\lambda$1666 &  39.254 &  39.271 &  38.710 &  38.747 &  38.262 &  38.336 \\
{[\ionf{O}{iii}]}    $\lambda$2321 &  38.542 &  38.552 &  38.061 &  38.076 &  37.676 &  37.697 \\
{[\ionf{O}{iii}]}    $\lambda$4363 &  39.142 &  39.152 &  38.661 &  38.676 &  38.276 &  38.296 \\
{[\ionf{O}{iii}]}    $\lambda$4959 &  40.564 &  40.568 &  40.162 &  40.164 &  39.834 &  39.834 \\
{[\ionf{O}{iii}]}   $\lambda$5007 &  41.038 &  41.042 &  40.636 &  40.639 &  40.308 &  40.308 \\
\noalign{\medskip}
\multicolumn{7}{c}{                          \ldots }\\
\noalign{\medskip}
\noalign{\medskip}
\hline
\end{tabular}
\end{center}
\begin{list}{}{}
\footnotesize
\item[\textbf{Note:}]Table \ref{tab:cloudy:output} is published in its entirety in the machine-readable format. A portion is shown here for guidance regarding its form and content.
\item[]Parameters in the example model are as follows: metallicity $Z/$Z$_{\odot}=0.5$, mass-loss rate $\dot{M}_{\rm sc} = 0.607 \times 10^{-2}$ M$_{\odot}$\,yr$^{-1}$, wind terminal speed $V_{\infty}= 457$ km\,s$^{-1}$, cluster radius $R_{\rm sc} = 1$ pc,
total stellar mass $M_{\star}= 2.05 \times 10^6$\,M$_{\odot}$, age $t=1$ Myr, ambient density $n_{\rm amb} = 100$ cm$^{-3}$, and ambient temperature $T_{\rm amb}$ calculated by \cloudy. 
\end{list}
\end{table}
 

\begin{itemize}
 \item[--] \verb|metal|: metallicity $\hat{Z} \equiv Z/$Z$_{\odot} = 1$, $0.5$, $0.25$, and $0.125$.

 \item[--] \verb|dMdt|: mass-loss rate $\dot{M}_{\rm sc} = 10^{-1}$, $10^{-2}$, $10^{-3}$, and $10^{-4} \times \hat{Z}^{0.72}$ M$_{\odot}$\,yr$^{-1}$.

 \item[--] \verb|Vinf|: wind terminal speed $V_{\infty}= 250$, $500$, and $1000 \times \hat{Z}^{0.13}$ km\,s$^{-1}$.

 \item[--] \verb|Rsc|: cluster radius $R_{\rm sc} = 1$ pc.

 \item[--] \verb|age|: current age $t = 1$ Myr.

 \item[--] \verb|Mstar|: total stellar mass $M_{\star}= 2.05 \times 10^6$\,M$_{\odot}$.

 \item[--] \verb|logLion|: ionizing luminosity $\log L_{\rm ion}$ (erg/s).
   
 \item[--] \verb|Namb|: ambient density $n_{\rm amb} = 1$, 10, $10^2$, and $10^3$ cm$^{-3}$.

 \item[--] \verb|Tamb|: mean ambient temperature $T_{\rm amb}$ determined by \cloudy.

 \item[--] \verb|Rmax|: maximum radius $R_{\rm max}$ (pc) for the surface brightness integration.

 \item[--] \verb|Raper|: aperture radius $R_{\rm aper}$ (pc) for the total luminosity integration.

 \item[--] \verb|Rshell|: shell outer radius $R_{\rm shell}$ (pc).
 
 \item[--] \verb|Rstr|: Str\"{o}mgren radius $R_{\rm str}$ (pc) determined by \cloudy.
 
 \item[--] \verb|Rbin|: bubble inner radius $R_{\rm b,in}$ (pc).
 
 \item[--] \verb|Rbout|: bubble outer radius $R_{\rm b,out}$ (pc) or shell inner radius.
 
 \item[--] \verb|Tbubble|: median temperature $T_{\rm bubble}$ of the hot bubble (region b in Figure~\ref{fig:temp:dens:profiles}).

 \item[--] \verb|Tadi|: median temperature $T_{\rm adi,med}$ of the expanding wind predicted by the adiabatic solutions (region a in Figure~\ref{fig:temp:dens:profiles}).
 
 \item[--] \verb|Twind|: median temperature $T_{\rm w,med}$ of the expanding wind calculated by \maihem with the radiative solutions (region a in Figure~\ref{fig:temp:dens:profiles}).
 
 \item[--] \verb|logUsp|: dimensionless ionization parameter $\log U_{\rm sph}$ in a spherical geometry calculated by \cloudy, defined as $U_{\rm sph} \equiv Q({\rm H}^0)/4 \pi R^2_{\rm str} n_{\rm H} c$, where $Q({\rm H}^0)$ is the total number of
hydrogen-ionizing photons emitted per second, $n_{\rm H}$ the hydrogen density, and $c$ the speed of light.
 
 \item[--] \verb|thin|: optically thin (1) or thick (0) model.
 
 \item[--] \verb|mode|: the cooling/heating radiative/adiabatic modes: 
 1 (AW: adiabatic wind), 
 2 (AB: adiabatic bubble), 3 (AP: adiabatic, pressure-confined), 
 4 (CC: catastrophic cooling), 5 (CB: catastrophic cooling bubble), 
 and 6 (CP: catastrophic cooling, pressure-confined). 
 
  \item[--] \verb|H_1_1216|, \verb|H_1_6563|, \ldots , \verb|Ar_5_7006|: integrated luminosities of the emission lines 
  Ly$\alpha$ $\lambda$1216 {\AA}, H$\alpha$ $\lambda$6563 {\AA}, \ldots , [\ionf{Ar}{v}] $\lambda$7006 {\AA},
  respectively. 
  
\end{itemize}

\end{appendix}

\vspace{10pt}

{ \small 
\begin{center}
\textbf{ORCID iDs}
\end{center}
\vspace{-5pt}

\noindent Ashkbiz~Danehkar \orcidauthor{0000-0003-4552-5997} \url{https://orcid.org/0000-0003-4552-5997}

\noindent M. S. Oey \orcidauthor{0000-0002-5808-1320} \url{https://orcid.org/0000-0002-5808-1320}

\noindent William J. Gray \orcidauthor{0000-0001-9014-3125} \url{https://orcid.org/0000-0001-9014-3125}

}



\end{document}